\documentclass[12pt]{article}

\usepackage{geometry,float}
\usepackage{graphicx}
\usepackage{amssymb,mathtools}
\usepackage{amsmath}
\usepackage{amsthm,bm,bbm}
\usepackage{color}
\usepackage{tabularx}
\usepackage{enumitem}
\usepackage[authoryear]{natbib}
\usepackage{dsfont}
\usepackage{algorithm}
\usepackage{algorithmic}
\usepackage{multirow}
\usepackage{JASA_manu}
\usepackage{url}
\usepackage{paralist}
\usepackage{hyperref}
\global\long\def\T{\T}
\renewenvironment{thebibliography}[1]{
\begin{oldthebibliography}{#1}
\setlength{\parskip}{0ex}
\setlength{\itemsep}{1ex}
\setlength{\baselineskip}{3.5ex}
}
{
\end{oldthebibliography}
}

\allowdisplaybreaks
\makeatletter
\newenvironment{breakablealgorithm}
{% \begin{breakablealgorithm}
\begin{center}
 \refstepcounter{algorithm}% New algorithm
 \hrule height.8pt depth0pt \kern2pt% \@fs@pre for \@fs@ruled »­Ïß
 \renewcommand{\caption}[2][\relax]{% Make a new \caption
   {\raggedright\textbf{\ALG@name~\thealgorithm} ##2\par}%
   \ifx\relax##1\relax % #1 is \relax
     \addcontentsline{loa}{algorithm}{\protect\numberline{\thealgorithm}##2}%
   \else % #1 is not \relax
     \addcontentsline{loa}{algorithm}{\protect\numberline{\thealgorithm}##1}%
   \fi
   \kern2pt\hrule\kern2pt
 }
}{% \end{breakablealgorithm}
 \kern2pt\hrule\relax% \@fs@post for \@fs@ruled »­Ïß
\end{center}
}
\makeatother

\newtheorem{thm}{Theorem}

\newtheorem{cor}{Corollary}
\newtheorem{lem}{Lemma}

\def\argmin{\mathop{\rm argmin}}
\def\var{\mathop{\rm Var}}
\def\cov{\mathop{\rm Cov}}
\def\v{{\varepsilon}}
\def\T{{\mathrm T}}
\def\mean{{\mathrm E}}
\def\idx{{j}}

\newcommand{\blind}{1}

\begin{document}
\def\spacingset#1{\renewcommand{\baselinestretch}%
{#1}\small\normalsize} \spacingset{1}

%%%%%%%%%%%%%%%%%%%%%%%%%%%%%%%%%%%%%%%%%%%%%%%%%%%%%%%%%%%%%%%%%%%%%%%%%%%%%%

\if1\blind
{
\title{\bf Online Estimation for Functional Data}
\author{Ying Yang and Fang Yao\thanks{
Fang Yao is the corresponding author, fyao@math.pku.edu.cn. This research is supported by National Natural Science Foundation of China Grants No.11931001 and 11871080, the LMAM, and the Key Laboratory of Mathematical Economics and Quantitative Finance (Peking University), Ministry of Education. }\hspace{.2cm}\\
Department of Probability and Statistics, School of Mathematical Sciences,\\
 Center for Statistical Science, Peking University, Beijing, China
}
\maketitle
} \fi

\if0\blind
{
\bigskip
\bigskip
\bigskip
\begin{center}
{\LARGE\bf Online Estimation for Functional Data}
\end{center}
\medskip
} \fi

\spacingset{1.45} % DON'T change the spacing!
\bigskip
\begin{abstract}

Functional data analysis has attracted considerable interest and is facing new challenges, one of which is the increasingly available data in a streaming manner. In this article we develop an  online nonparametric method to dynamically update the estimates of mean and covariance functions for functional data. The kernel-type estimates can be decomposed into two sufficient statistics depending on the data-driven bandwidths. We propose to approximate the future optimal bandwidths by a sequence of dynamically changing candidates and combine the corresponding statistics across blocks to form the updated estimation. The proposed online method is easy to compute based on the stored sufficient statistics and the current data block. We derive the asymptotic normality and, more importantly, the relative efficiency lower bounds of the online estimates of mean and covariance functions. This provides insight into the relationship between estimation accuracy and computational cost driven by the length of candidate bandwidth sequence. Simulations and real data examples are provided to support such findings.

\end{abstract}

\noindent%
\vfill

\newpage
% \linenumbers
\section{Introduction}\label{sec:introduction}

\begin{figure}[htbp]
\centering
\includegraphics[width = 5in]{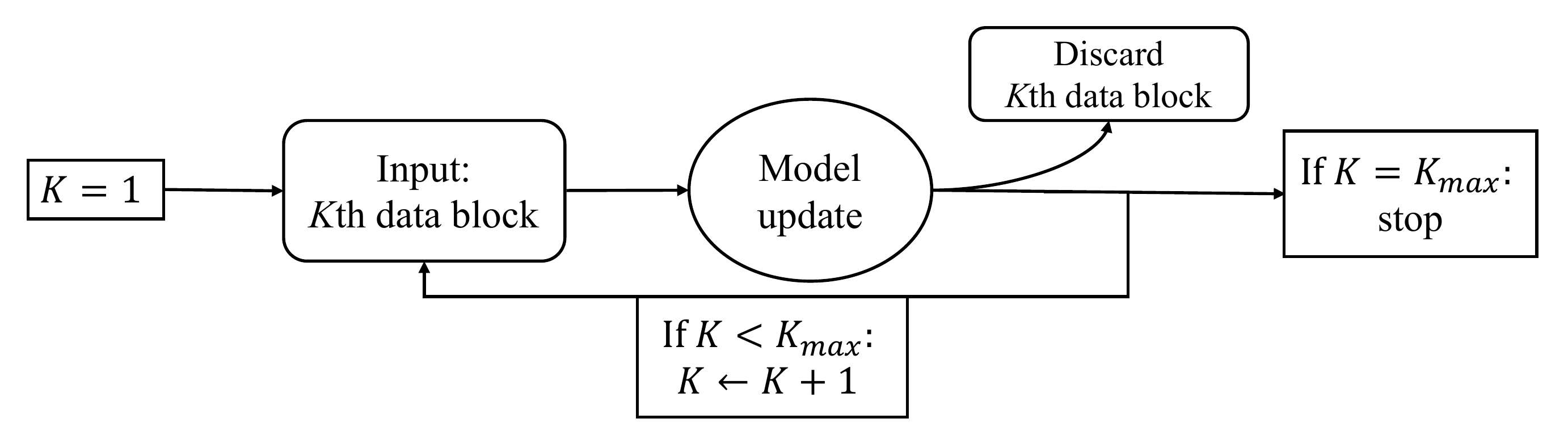}
\caption{\label{fig:online} Online estimation scheme. }
\end{figure}

Modern technology has promoted the prevalence of functional data in many fields, such as biomedical studies, engineering, social sciences and so on. A fundamental problem in functional data analysis is the estimation of mean and covariance functions, which sets stage for subsequent analyses such as functional principal component analysis and functional regression. 
Technological advancements have improved the speed and volume of data acquisition and brought new challenges to model such data in a streaming manner using {\em online} approaches, which updates the model estimation with only the current available data and avoid storing the previous data, as illustrated in Figure \ref{fig:online}. Such online methods not only provide a way to analyze the out-of-memory data but also allow real-time output of results, and hence are practically useful.
It has been studied extensively in machine learning and statistics. \cite{Langford2009Sparse}, \cite{Duchi2009Efficient}, \cite{Xiao2010Dual} and \cite{Dekel2012Optimal} discussed stochastic gradient descent problems with explicit regularizations. \cite{Schifano2016Online} proposed an online version of predictive residual test for linear model and improved the aggregated estimating equation of \cite{lin2011aggregated} for generalized linear model. Extensions of linear and quadratic discriminant analyses have also been studied, see \cite{hiraoka2000convergence}, \cite{kim2007incremental} and \cite{pang2005incremental}. Following the custom of these papers, we refer to the classical approaches using the full data as {\em batch} methods. 
To extend functional data analysis to the online context, the primary interest is to effectively and efficiently estimate the mean and covariance functions under various sampling schemes. Suppose that we have $n$ subjects which are observed at $m_i$ time points within the domain $\mathcal{T}$ for the $i$th individual, denoted by $T_i=(T_{i1},\ldots,T_{im_i}), i=1,\ldots,n$. In this work, we focus on the case that each subject is observed at possibly different times, i.e., $T_i$ are randomly distributed, referred to as {\em independent} design. This is distinguished from the {\em common} design where all subjects are observed on a common set of times, as discussed in \cite{cai_optimal_2011}. From the online perspective, the latter is easier for it may take sample mean and sample covariance as sufficient statistics and store historical data in a linear fashion. For independent design, there are typically two types of estimation methods: pre-smoothing and pooling methods. The former is to smooth each individual curve and take the cross-sectional average to form the estimates, which is additive over subjects and can be extended to online settings straightforwardly. However, pre-smoothing technique is only applicable to sufficiently dense data and requires $m_i\gg n^{5/4}$, i.e., $m_i/ n^{5/4}\rightarrow C\in(0,\infty]$,
% $m_i\gg n^{5/4}$, where $a\gg b$ means $a/b\rightarrow C\in(0,\infty]$ as $a,b\rightarrow\infty$,
to attain the root-$n$ convergence rate for the mean and covariance estimation when tuning parameter (e.g., bandwidth) is chosen optimally for each subject, see \cite{zhang2007inference} and \cite{kong_partially_2016}. By contrast, pooling method can be adopted to both sparse and dense data. It borrows information from neighboring data and pools all subjects to estimate the mean and covariance functions, which is typically achieved by nonparametric methods including splines \citep{rice1991estimating, sugar2000principal,Peng2009Consistency} and local polynomials \citep{yao2005sparse,yao2005functional,li2010uniform}, the later only requires $m_i\gg n^{1/4}$ to attain the root-$n$ convergence rate \citep{zhang2016from}. 

In this work, we focus on the pooling method using local polynomial smoothing which is still challenging to extend to online estimation with theoretical guarantees. Here we take advantage of the extensive theory of local polynomial regression. Specifically, such kernel-based estimators have the form of locally weighted least squares and can be decomposed into two sufficient statistics that are additive on data depending on the bandwidth. The major difficulty is that the optimal bandwidth changes over data collection and the sufficient statistics vary accordingly. For instance, \cite{zhou2003m}, \cite{Kristan2010Online} and \cite{kong2019efficiency} studied online local polynomial estimation, but did not update the bandwidths when accessing the stored statistics from previous data, thus lost the optimality to some extent.	To overcome this, our key proposal is to generate a dynamic sequence of candidate bandwidths that contain the current optimal bandwidth as well as smaller values to approximate the future optimal bandwidths. The corresponding statistics based on the current data block with these dynamically updated candidate bandwidths are computed, which contain information only from one data block and are referred to as ``sub-statistics''. The stored sufficient statistics are then updated by aggregating with the current sub-statistics. Consequently, we are able to update the estimate based on not only the current data block but also historical data, while only storing one set of dynamically updated sufficient statistics.

We employ the proposed online mean and covariance estimation to both dense and sparse functional data, which are also accessible for theoretical analysis. We derive the asymptotic normality of the online mean and covariance estimates from sparse to dense designs that is shown to coincide with the phase transition regime in \cite{zhang2016from}, even when the estimated mean is used in the covariance estimation. More importantly, when compared to the classical results based on full data, our online estimates have explicit lower bounds of the relative efficiency attained by the dynamically updated bandwidth sequences. These lower bounds indicate that the efficiency is proportional to the length of candidate bandwidth sequence, which suggests computing time and memory needed in a linear manner, especially for similar block sizes. Therefore we can make an informed choice on the trade-off between estimation accuracy and computational cost according to the real problems at hand. 

The rest of the paper is organized as follows. In Section \ref{sec:mean and cov}, we briefly describe the classical estimation of the mean and covariance functions for functional data analysis. In Section \ref{sec:method}, we elucidate the proposed online method for the mean and covariance estimation using dynamic candidate bandwidths. In Section \ref{sec:th}, we investigate theoretical properties of the proposed method and establish the lower bounds of the relative efficiency for the online estimators. Simulations and real data applications are presented in Section \ref{sec:simu} and \ref{sec:real data} to validate our theoretical findings. Technical assumptions and bandwidth selection are delineated in the Appendix, and proofs of theoretical results are collected in an online Supplementary Material for space economy. 

\section{Mean and Covariance Estimation}
\label{sec:mean and cov}
We first review the classical estimation of the mean and covariance functions for functional data analysis. Let $\{X(t):t\in \mathcal{T}\}$ be an $L^2$ stochastic process on an interval $\mathcal{T}$ that can be expressed as
\begin{equation*}
X(t)=\mu(t)+\Phi(t),
\end{equation*}
where $\mu(t)=\mean\{X(t)\}$ is the mean function, $\Phi(t)$ is the stochastic part of $X(t)$ with $\mean\Phi(t)=0$ and $\cov\{\Phi(s),\Phi(t)\}=\gamma(s,t)$ for all $s,t\in \mathcal{T}$, which implies that $\cov\{X(s),X(t)\}=\gamma(s,t)$. 
In the sequel, $\mathcal{T}$ is assumed to be $[0,1]$ for simplicity.

Denote the current block as $K$ which increases to $K_{max}$ that tends to $\infty$ as data accumulate. Suppose that we have observed $n_k$ subjects in the $k$th data block for $k=1,\ldots,K$, among which the $i$th subject has $m_{ki}$ measurements at possibly irregular time points, $T_{kij}$, $i=1,\ldots,n_k$, $j=1,\ldots,m_{ki}$. In this work, we focus on the setting where only the number of subjects are increasing as data accumulate. %Specifically, let $(k,i)$ represent the $i$th subject in the $k$th block, then for each $(k,i)$, its measurements are contained only in the $k$th block.   
This setting is different from the online time series analysis, where the measurements of one or a few sequences are increasing, as discussed in \cite{bacher_online_2009}, \cite{2013Online}, \cite{yang_online_2019} and \cite{richard_online_2009}. They usually considered, e.g., autoregressive models with time-invariant parameters, hence the accumulation of measurements can improve the estimation.  By contrast, in functional data analysis, the trajectory of each subject is viewed as a random function, while the mean/covariance functions are modeled nonparametrically with information accumulated across subjects to improve estimation. We further remark that, if measurements for each subject are also collected in a streaming fashion, it raises the challenge due to dependence between newly arriving measurements and the previously discarded ones from the same subjects when considering nonparametric modeling of the mean/covariance function. One possibility is to incorporate some parametric structures such as those in time series models, coupled with functional data methods. This leads to a different line of research that can be a topic of future investigation. 

Suppose that observations are contaminated with noises following
\begin{equation}
\label{model fda}
Y_{kij}=X_{ki}(T_{kij})+\v_{kij}=\mu(T_{kij})+\Phi_{ki}(T_{kij})+\v_{kij},
\end{equation}
where $\v_{kij}$ are independently and identically distributed (i.i.d) with $\mean(\v_{kij})=0$ and $\var(\v_{kij})=\sigma^2$. For notations, the subscript ``$k$'' or ``$K$'' denotes statistics based on only the $k$th or $K$th data block. The superscript ``$(k)$'' or ``$(K)$'' denotes statistics calculated either using the full data up to $k$ or $K$ by classical batch method (denoted by `` $\widehat{}$ ''), or using only the $k$- or $K$-th data block and stored statistics by online method (denoted by `` $\widetilde{}$ ''), both of which contain information of data blocks up to $k$ or $K$.  Recall that \citet{zhang2016from} discussed two weighting schemes, equal weight per observation/subject, which has no material difference when coupled with online estimation except for their expressions. Thus we focus on the case of equal weight per observation for conciseness in the sequel. 

Denote $W_h(\cdot)=W(\cdot/h)/h$, where $W(\cdot)$ is the kernel function and $h$ is the bandwidth. Classical mean estimate based on local linear smoother at block $K$ is $\widehat\mu^{(K)}(t)=\widehat\alpha_0$, where
\begin{equation*}
% \label{eq:estimate mu}
(\widehat\alpha_0,\widehat\alpha_1)=\argmin\limits_{\alpha_0,\alpha_1}\sum_{k=1}^K\sum_{i=1}^{n_k}\sum_{j=1}^{m_{ki}}\{Y_{kij}-\alpha_0-\alpha_1(T_{kij}-t)\}^2W_{\widehat{h}_{\mu}^{(K)}}(T_{kij}-t),
\end{equation*} 
and $\widehat{h}_{\mu}^{(K)}$ is the bandwidth selected by the batch method using the full data up to block $K$.  Let $e_2 = (1,0)^\T$ and $T_{kij}(t)=(1,T_{kij}-t)^\T$, and the solution can be explicitly written as
\begin{equation}
\label{eq:solution mu}
\widehat{\mu}^{(K)}(t) = e_2^\T\left\{\sum_{k=1}^KP_k\big(t;\widehat{h}_{\mu}^{(K)}\big)\right\}^{-1}\left\{\sum_{k=1}^Kq_k\big(t;\widehat{h}_{\mu}^{(K)}\big)\right\},
\end{equation}
where $\{P_k,q_k\}$ only depends on the $k$th block given $\widehat{h}_{\mu}^{(K)}$ by
\begin{align} 
&P_k(t;\widehat{h}_{\mu}^{(K)})=\sum_{i=1}^{n_k}\sum_{j=1}^{m_{ki}}W_{\widehat{h}_{\mu}^{(K)}}(T_{kij}-t)T_{kij}(t)T_{kij}(t)^\T
,\nonumber\\
&q_k(t;\widehat{h}_{\mu}^{(K)})=\sum_{i=1}^{n_k}\sum_{j=1}^{m_{ki}}W_{\widehat{h}_{\mu}^{(K)}}(T_{kij}-t)T_{kij}(t)Y_{kij}.\nonumber
\end{align}

To estimate $\gamma$, let $m_{ki}\geq2$ for no contribution to covariance estimation otherwise. Let $\widehat{C}_{ki,j_1,j_2}^{(K)}$ be the raw covariance based on $\widehat{\mu}^{(K)}$ in \eqref{eq:solution mu}, i.e. 
\begin{equation} 
\label{eq:Cbatch}
\widehat{C}_{ki,j_1,j_2}^{(K)}=\{Y_{kij_1}-\widehat{\mu}^{(K)}(T_{kij_1})\}\{Y_{kij_2}-\widehat{\mu}^{(K)}(T_{kij_2})\},
\end{equation}
where $1\le j_1\neq j_2\le m_{ki}$. Let $\widehat{h}_{\gamma}^{(K)}\in\mathbb{R}$ be the batch bandwidth up to block $K$ for covariance estimation, and the estimate of $\gamma$ is give by $\widehat\gamma^{(K)}(s,t)=\widehat\tau_0$, where
\begin{eqnarray*}
(\widehat\tau_0,\widehat\tau_1,\widehat\tau_2)=\argmin\limits_{\tau_0,\tau_1,\tau_2}
\sum_{k=1}^K\sum_{i=1}^{n_k}\qquad\sum_{\mathclap{1\le j_1\neq j_2\le m_{ki}}}
\big\{\widehat{C}_{ki,j_1,j_2}^{(K)}-\tau_0-\tau_1(T_{kij_1}-s)-\tau_2(T_{kij_1}-t)\big\}^2\\ 
\times W_{\widehat{h}_{\gamma}^{(K)}}(T_{kij_2}-s)W_{\widehat{h}_{\gamma}^{(K)}}(T_{kij_2}-t).
\end{eqnarray*} 
Let $e_3 = (1,0,0)^\T$ and $T_{ki,j_1,j_2}(s,t)=(1,T_{kij_1}-s,T_{kij_2}-t)^\T$, and the explicit solution is 
\begin{equation}
\label{eq:solution gamma}
\widehat{\gamma}^{(K)}(s,t) = 
e_3^\T\left\{\sum_{k=1}^KP_k\big(s,t;\widehat{h}_{\gamma}^{(K)}\big)\right\}^{-1}\left\{\sum_{k=1}^Kq_k\big(s,t;\widehat{h}_{\gamma}^{(K)}\big)\right\},
\end{equation}
where
\begin{align} 
&P_k(s,t;\widehat{h}_{\gamma}^{(K)})=\sum_{i=1}^{n_k}\qquad\sum_{\mathclap{1\le j_1\neq j_2\le m_{ki}}} W_{\widehat{h}_{\gamma}^{(K)}}(T_{kij_1}-s)W_{\widehat{h}_{\gamma}^{(K)}}(T_{kij_2}-t)T_{ki,j_1,j_2}(s,t)T_{ki,j_1,j_2}(s,t)^\T,\nonumber\\
&q_k(s,t;\widehat{h}_{\gamma}^{(K)})=\sum_{i=1}^{n_k}\qquad\sum_{\mathclap{1\le j_1\neq j_2\le m_{ki}}} W_{\widehat{h}_{\gamma}^{(K)}}(T_{kij_1}-s)W_{\widehat{h}_{\gamma}^{(K)}}(T_{kij_2}-t)T_{ki,j_1,j_2}(s,t)\widehat{C}_{ki,j_1,j_2}.\nonumber
\end{align}

\section{Proposed Online Method}
\label{sec:method}

In this section, we give a detailed description of the online estimation for mean function $\mu(t)$, and the covariance estimation can be derived similarly. It can be seen from \eqref{eq:solution mu} that $\sum_{k=1}^KP_k\big(t;\widehat{h}_\mu^{(K)}\big)$ and $\sum_{k=1}^Kq_k\big(t;\widehat{h}_\mu^{(K)}\big)$ are a pair of precise sufficient statistics. Given $\widehat{h}_\mu^{(K)}$, we only need to store $\sum_{k=1}^KP_k\big(t;\widehat{h}_\mu^{(K)}\big)$ and $\sum_{k=1}^Kq_k\big(t;\widehat{h}_\mu^{(K)}\big)$ in computer memory instead of the entire $K$ data blocks. However, it is noteworthy that $\widehat{h}_\mu^{(K)}$ takes different values as block $K$ varies, e.g.,	$K\rightarrow\infty$ in streaming problems. This data-driven feature makes the online local polynomial estimation rather difficult, as storing all $\big\{\sum_{k=1}^KP_k\big(t;\widehat{h}_\mu^{(K)}\big),\sum_{k=1}^Kq_k\big(t;\widehat{h}_\mu^{(K)}\big)\big\}_{K=1}^{K_{max}}$ violates the virtue of online method. For example, \cite{kong2019efficiency} did not re-calculate the stored statistics with the updated bandwidth. This is a general issue for online statistical estimation involving data-driven parameters or structures and partially explains lack of proper online algorithms for local polynomial estimation that is arguably one of the most popular nonparametric regression methods. 

To overcome this obstacle, we propose to generate a dynamic sequence at block $k$ ($k\leq K$) consisting of $L$ candidate bandwidths, denoted by $\{\eta_{\mu,l}^{(k)}\}_{l=1}^L$. The key idea is to use appropriately selected $\eta_{\mu,l}^{(k)}$ as a surrogate of future optimal bandwidths $h_{\mu,*}^{(K)}$ for $k\le K$. 
{For the flow of presentation, the expressions of $h_{\mu,*}^{(K)}$ and the corresponding online estimator $\widetilde{h}_\mu^{(k)}$ as well as the optimal form for $\{\eta^{(k)}_{\mu,l}\}_{l=1}^L$ are deferred to \eqref{opt hs}, \eqref{online h} and \eqref{form of eta fda} of Section \ref{sec:th}.}
% {\color{red}Let $\widetilde{h}_\mu^{(k)}$ be the online estimator of $h_{\mu,*}^{(k)}$ given explicitly in \eqref{online h} of Section \ref{sec:th}.} 
We prove in Theorem \ref{THM:BAND} that $\widetilde{h}_\mu^{(k)}$ converges to $h_{\mu,*}^{(k)}$ when $k\rightarrow\infty$, and propose to select from $\{\eta_{\mu,l}^{(k)}\}_{l=1}^L$ to approximate $\widetilde{h}_{\mu}^{(K)}$ empirically. 
Note that $\widetilde{h}_\mu^{(k)}$ decreases with respect to $k$, i.e., $\widetilde{h}_{\mu}^{(k)}\ge\widetilde{h}_{\mu}^{(K)}$ for $k\le K$, we set $\widetilde{h}_\mu^{(k)}=\eta_{\mu,1}^{(k)}> \eta_{\mu,2}^{(k)}> \ldots> \eta_{\mu,L}^{(k)}$. 
% {\color{red}The optimal form for $\{\eta^{(k)}_{\mu,l}\}_{l=1}^L$ are given in \eqref{form of eta fda} of Section \ref{sec:th}.} 
At every block $k$, we compute $L$ pairs of statistics with candidate bandwidths $\{\eta_{\mu,l}^{(k)}:l=1,2,\ldots,L\}$ based on the $k$th data block. This gives a sequence of statistics $\{P_k\big(t;\eta_{\mu,l}^{(k)}\big),q_k\big(t;\eta_{\mu,l}^{(k)}\big)\}_{l=1}^L$ referred to as ``sub-statistics'' as they contain only information of the $k$th block \big(the definition of $P_k(t;\cdot),q_k(t;\cdot)$ is given in \eqref{eq:solution mu}\big). To combine information across blocks, we introduce the so-called ``pseudo-sufficient statistics'' $\{\widetilde{P}_{\mu,l}^{(K)},\widetilde{q}_{\mu,l}^{(K)}\}_{l=1}^L$, whose update depends on only the sub-statistics of the current data block and the stored pseudo-sufficient statistics of previous blocks. %We now describe the combination rule in detail.
Specifically, 
% let $\idx_l^{(K)}$ be the index of the one of previous pseudo-sufficient statistics $\{\widetilde{P}_{\mu,l}^{(K-1)},\widetilde{q}_{\mu,l}^{(K-1)}\}_{l=1}^L$ to combine with the current $l$th sub-statistics $\{P_K\big(t;\eta_{\mu,l}^{(K)}\big),q_K\big(t;\eta_{\mu,l}^{(K)}\big)\}$, then the pseudo-sufficient statistics are updated as follows,
\begin{eqnarray}
\label{eq:update Pq}
\widetilde{P}_{\mu,l}^{(K)}(t)=P_K\big(t;\eta_{\mu,l}^{(K)}\big)+\widetilde{P}_{\mu,\idx_l^{(K)}}^{(K-1)}(t),\quad
\widetilde{q}_{\mu,l}^{(K)}(t)=q_K\big(t;\eta_{\mu,l}^{(K)}\big)+\widetilde{q}_{\mu,\idx_l^{(K)}}^{(K-1)}(t)
\end{eqnarray}
with initialization $\widetilde{P}_{\mu,l}^{(0)}(t)=0,\ \widetilde{q}_{\mu,l}^{(0)}(t)=0$, where the definition of $\idx_l^{(K)}$ would be presented later.
The formula \eqref{eq:update Pq} indicates that $\{\widetilde{P}_{\mu,1}^{(K)},\widetilde{q}_{\mu,1}^{(K)}\}$ are of form $\big\{\sum_{k=1}^{K}P_k(t;\cdot),\sum_{k=1}^{K}q_k(t;\cdot)\big\}$ which are calculated at different candidates across blocks, and hence the motivation is to select $\idx_l^{(K)}$ appropriately to make these candidates close to $\widetilde{h}_\mu^{(K)}$.
% The key problem is to select $\idx_l^{(K)}$ appropriately.
% We also need carefully select ``$\idx_l^{(K)}$'' for $l=2,\ldots,L$ to construct a sequence of statistics for future estimation with the intuition to make the candidates used in $\{\widetilde{P}_{\mu,l}^{(K)},\widetilde{q}_{\mu,l}^{(K)}\}$ close to $\eta_{\mu,l}^{(K)}$.  
 
\begin{figure}[htbp]
\centering
\includegraphics[width = 6in, height = 2.5in]{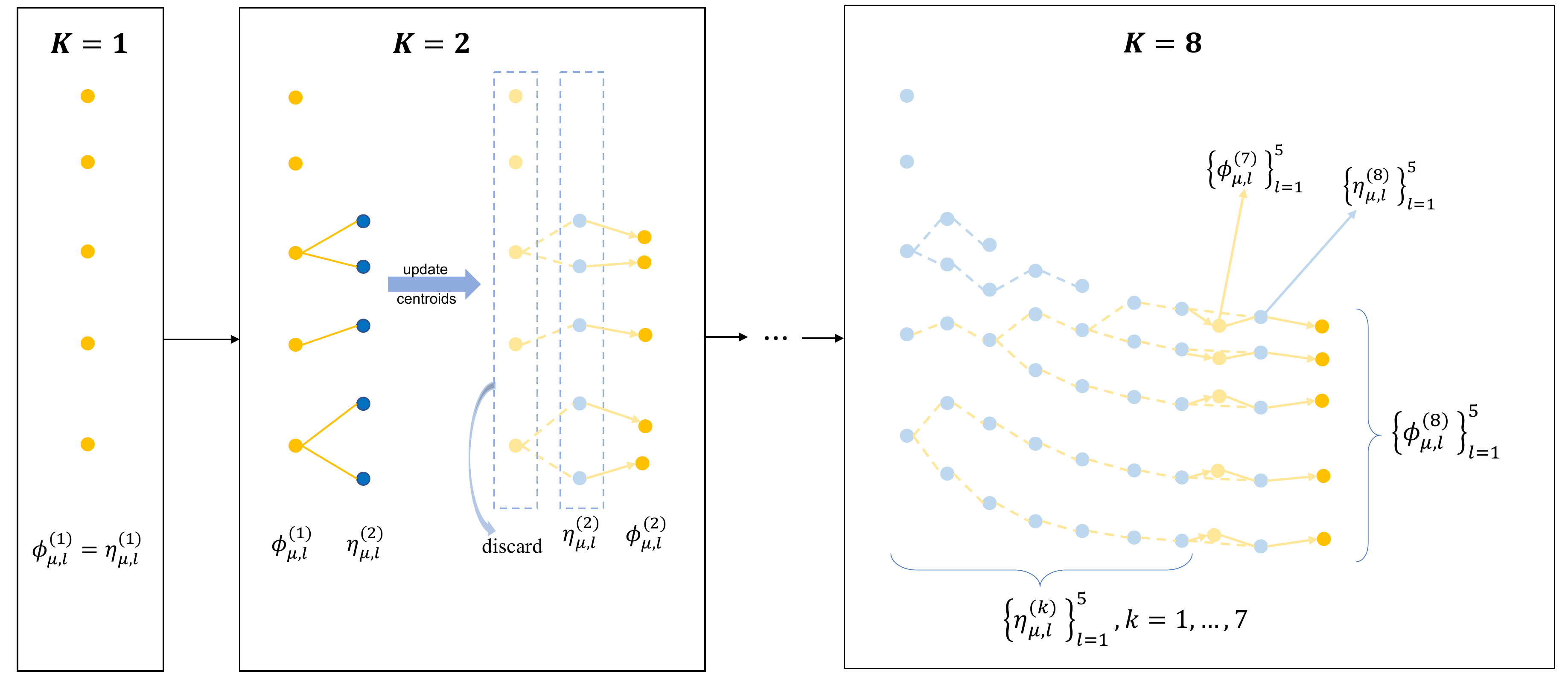}
\caption{\small\label{fig:sketch0} 
A sketch for the update procedure with $L = 5$ and $K = 1,2,8$. 
When $K=1$ (the left panel), $\omega_{\mu,1}^{(1)}=1$ and it can be obtained from \eqref{eq:centroids} and \eqref{eq:update Pq} that $\phi_{\mu,l}^{(1)}=\eta_{\mu,l}^{(1)}$ and $\widetilde{P}_{\mu,l}^{(1)}(t)=P_1\big(t;\eta_{\mu,l}^{(1)}\big),\widetilde{q}_{\mu,l}^{(1)}(t)=P_1\big(t;\eta_{\mu,l}^{(1)}\big)$ for $1\le l \le L$. 
When $K=2$, $\{\eta_{\mu,l}^{(2)}\}_{l=1}^L$ are generated, and for each $l$, $\idx_l^{(2)}$ selects the one closest to $\eta_{\mu,l}^{(2)}$ from $\{\phi_{\mu,i}^{(1)}\}_{i=1}^L$, then the corresponding statistics are aggregated as illustrated by the solid lines in the middle panel. Once $\idx_l^{(2)}$ are determined, we update $\{\phi_{\mu,l}^{(2)}\}_{l=1}^L$ and $\big\{\widetilde{P}_{\mu,l}^{(2)},\widetilde{q}_{\mu,l}^{(2)}\big\}_{l=1}^L$ according to \eqref{eq:centroids} and \eqref{eq:update Pq}, and $\{\idx_l^{(2)},\phi_{\mu,l}^{(1)},\eta_{\mu,l}^{(2)}\}_{l=1}^L$ can be discarded. 
For $K=8$, we select $\idx_l^{(8)}$ an update $\{\phi_{\mu,i}^{(8)}\}_{i=1}^L$ accordingly, and the procedure continues. By \eqref{eq:update Pq}, $\{\widetilde{P}_{\mu,l}^{(8)}(t),\widetilde{q}_{\mu,l}^{(8)}(t)\}_{l=1}^L$ are comprised of the sub-statistics of blockwise candidates connected by the dashed lines for the first 8 blocks in the right panel. 
}
\end{figure}

\begin{figure}[htbp]
\centering
\includegraphics[width = 6.5in]{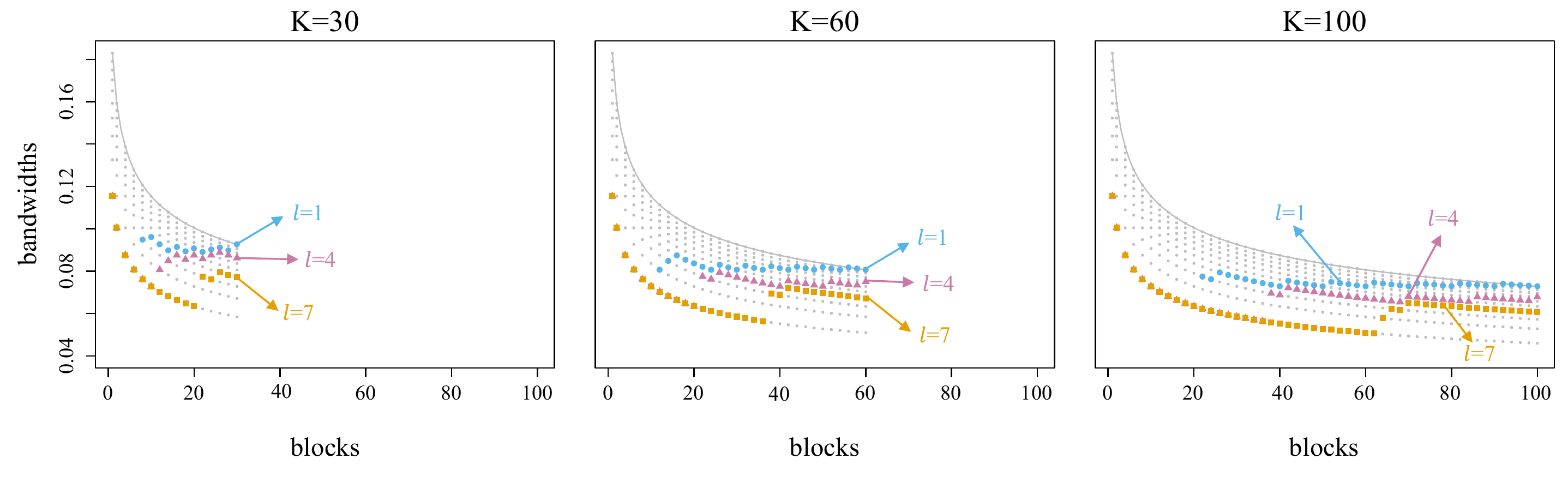}
\caption{\small\label{fig:sketch}	The solid line by connecting the largest values at each block corresponds to the online bandwidths $\widetilde{h}_\mu^{(k)},k=1,2,\ldots,K$. For illustration, we set $\widetilde{h}_\mu^{(k)}=CS_{k,1}^{-1/5}$ with $C=0.4, S_{k,1}=50k$ for $k=1,\ldots,100$. The vertical dot series represent the dynamic candidate sequences $\{\eta_{\mu, l}^{(k)}\}_{l=1}^L$, $k=1,2,\ldots,K$, which are calculated by \eqref{form of eta fda} given $\widetilde{h}_\mu^{(k)}$.	The bold points are the bandwidths used in $\big\{\widetilde{P}_{\mu,l}^{(K)},\widetilde{q}_{\mu,l}^{(K)}\big\}$ whose weighted averages are the centroids $\phi_{\mu,l}^{(K)}$ with $l=1,4,7$, respectively.  }
\end{figure}

Recall that there are $n_k$ subjects in the $k$th block and the $i$th one has $m_{ki}$ measurements. Define
\begin{equation}\label{n}
s_{k,j}=\sum_{i=1}^{n_k}m_{ki}(m_{ki}-1)\cdots(m_{ki}-j+1),\ S_{K,j}=\sum_{k=1}^Ks_{k,j},\ j\ge1,
\end{equation}
i.e., $s_{k,1},s_{k,2}$ are the sub-total numbers of observations of the $k$th block for estimating $\mu$ and $\gamma$, and $S_{K,1},S_{K,2}$ are the total numbers of observations up to block $K$, respectively. 
Let $\omega_{\mu,k}^{(K)}=s_{k,1}/S_{K,1}$ be the weight with respect to numbers of observations. 
To update $\idx_l^{(K)}$ in an online fashion, we define the centroids $\{\phi_{\mu,l}^{(K)}\}_{l=1}^L$ as the weighted average of all previous candidate bandwidths whose sub-statistics are aggregated, i.e.,
\begin{equation}
\label{eq:centroids}
\phi_{\mu,l}^{(0)}=0,\ \phi_{\mu,l}^{(K)}=(1-\omega_{\mu,K}^{(K)})\phi_{\mu,\idx_l^{(K)}}^{(K-1)}+\omega_{\mu,K}^{(K)}\eta_{\mu,l}^{(K)}.
\end{equation}
We now give the definition of $\idx_l^{(K)}$.
For $1\le l\le L$ and $1\le K\le K_{max}$, we propose to select $\idx_l^{(K)}$ to be the index of the closest one of $\{\phi_{\mu,l}^{(K-1)}\}_{l=1}^L$ (available at block $K$) to $\eta_{\mu,l}^{(K)}$, i.e.,
\begin{equation} 
\label{eq:I(l)}
\idx_l^{(K)}={\rm argmin}_{i\in\{1,2,\ldots,L\}}\vert\eta_{\mu,l}^{(K)}-\phi_{\mu,i}^{(K-1)}\vert.
\end{equation}
% The formula \eqref{eq:centroids} indicates that $\phi_{\mu,l}^{(K)}$ is the weighted average of the combined candidate bandwidths for $1\le l\le L$. 
The selection of $\idx_l^{(K)}$ in \eqref{eq:I(l)} guarantees that $\phi_{\mu,\idx_l^{(K)}}^{(K-1)}$ is close to $\eta_{\mu,l}^{(K)}$, which makes $\phi_{\mu,l}^{(K)}$ in \eqref{eq:centroids} close to $\eta_{\mu,l}^{(K)}$ as well. Combined with the definition of $\phi_{\mu,l}^{(K)}$, we conclude that the combined candidate bandwidths to produce $\phi_{\mu,l}^{(K)}$ are close to $\eta_{\mu,l}^{(K)}$ on average. Note that $\eta_{\mu,1}^{(K)}=\widetilde{h}_\mu^{(K)}$, the regression estimate at block $K$ is given by
\begin{equation}
\label{online mu}
\widetilde{\mu}^{(K)}(t)=e_{2}^\T\left\{\widetilde{P}_{\mu,1}^{(K)}(t)\right\}^{-1}\widetilde{q}_{\mu,1}^{(K)}(t).
\end{equation}   
Since the definitions of $\idx_l^{(K)}$ and $\{\phi_{\mu,l}^{(K)}\}_{l=1}^L$ are entangled with each other, we take $K=1, 2, 8$ and $L=5$ for example in Figure \ref{fig:sketch0} to illustrate the update procedure. We also display the candidates that produce $\phi_{\mu,i}^{(K)}$ for $i=1,4,7$ with $L=10$ and $K=30,60,100$ in Figure \ref{fig:sketch}. Note that one does not need to store the previous candidates or the pseudo-bandwidths in practice. 
% For each $l$, the selection of $\idx_l^{(K)}$ guarantees that $\phi_{\mu,l}^{(K)}$ is close to $\eta_{\mu,l}^{(K)}$, and hence $\big\{\widetilde{P}_{\mu,l}^{(K)},\widetilde{q}_{\mu,l}^{(K)}\big\}$ employ the bandwidths closest to $\eta_{\mu,l}^{(K)}$ 
% not only for the $(K-1)$th block but also 
% for all previous blocks (on average).

% \textcolor{red}(This description seems inconsistent with the figures!) 
To better appreciate \eqref{online mu}, let $\widetilde{\eta}_{\mu,k}^{(K)}$ denote the candidate bandwidth of the $k$th block used in $\widetilde{P}_{\mu,1}^{(K)}$ and $\widetilde{q}_{\mu,1}^{(K)}$, which correspond to the bold dots of $l=1$ in Figure \ref{fig:sketch} ($K=30,60,100$). We refer to the dynamic sequence $\{\widetilde{\eta}_{\mu,k}^{(K)}\}_{k=1}^K$ as the ``pseudo-bandwidth'' chain at block $K$ and formulate an equivalent expression of $\widetilde{\mu}^{(K)}(t)$ in \eqref{online mu}: 
\begin{equation}
\label{eq:online mu 2}
\widetilde{\mu}^{(K)}(t)=e_2^\T\left\{\sum_{k=1}^KP_k\big(t;\widetilde{\eta}_{\mu,k}^{(K)}\big)\right\}^{-1}\left\{\sum_{k=1}^Kq_k\big(t;\widetilde{\eta}_{\mu,k}^{(K)}\big)\right\},
\end{equation}
where $P_k(t;\cdot),q_k(t;\cdot)$ are defined as in \eqref{eq:solution mu}. Compared \eqref{eq:online mu 2} to the batch estimate \eqref{eq:solution mu}, we are using $\widetilde{\eta}_{\mu,k}^{(K)}, k=1,2,\ldots,K$ to approximate $\widehat{h}_\mu^{(K)}$ which converges to the optimal bandwidth  $h_{\mu,*}^{(K)}$. As mentioned, the definition of centroids and the combination rule for statistics guarantee that pseudo-bandwidth chain $\{\widetilde{\eta}_{\mu,k}^{(K)}\}_{k=1}^K$ is close to $\eta_{\mu,1}^{(K)}=\widetilde{h}_\mu^{(K)}$ on average, as shown in Figure \ref{fig:sketch}. Hence the online estimation using the pseudo-bandwidths shall perform well as long as $\widetilde{h}_\mu^{(K)}$ converges to $h_{\mu,*}^{(K)}$.
% , where $h_{\mu,*}^{(K)}$ is the optimal bandwidth at block $K$ that is given explicitly in \eqref{opt hs} of Section \ref{sec:th}. 
An implementation algorithm is given at the end of this section. We mention that $\{\eta_{\mu,l}^{(K)}\}_{l=1}^L$ and $\{\idx_l^{(K)}\}_{l=1}^L$ are recalculated at each $K$, and only the newest $\{\phi_{\mu,l}^{(K)}\}_{l=1}^L$ and $\big\{\widetilde{P}_{\mu,l}^{(K)},\widetilde{q}_{\mu,l}^{(K)}\big\}_{l=1}^L$ are stored in memory during the procedure. Hence the algorithm is computational efficient.

For estimating the covariance function $\gamma(s,t)$, similar to \eqref{eq:Cbatch}, define the ``raw'' covariance observations for each block based on the online mean estimate as
\begin{equation}
\widetilde{C}_{ki,j_1,j_2}^{(k)}=\{Y_{kij_1}-\widetilde{\mu}^{(k)}(T_{kij_1})\}\{Y_{kij_2}-\widetilde{\mu}^{(k)}(T_{kij_2})\},
\end{equation}
where $1\le j_1\neq j_2\le m_{ki},\ i=1,\ldots,n_k$.  
Let $\widetilde{h}_\gamma^{(K)}$ be the online bandwidth given explicitly in \eqref{online h} of Section \ref{sec:th} and $\{\eta_{\gamma,l}^{(K)}\}_{l=1}^L$ be the dynamic candidate bandwidths at block $K$ satisfying $\widetilde{h}_\gamma^{(K)}=\eta_{\gamma,1}^{(K)}>\ldots\eta_{\gamma,L}^{(K)}$. 
Then by setting $\omega_{\gamma,k}^{(K)}=s_{k,2}/S_{K,2}$ and substituting $\eta_{\mu,l}^{(K)}$ with $\eta_{\gamma,l}^{(K)}$ in \eqref{eq:update Pq}, \eqref{eq:I(l)} and \eqref{eq:centroids}, we obtain the dynamic centroids $\{\phi_{\gamma,l}^{(k)}\}_{l=1}^L$ and pseudo-sufficient statistics $\{\widetilde{P}_{\gamma,l}^{(K)},\widetilde{q}_{\gamma,l}^{(K)}\}_{l=1}^L$ which are of form $\{\sum_{k=1}^KP_k(s,t;\cdot),\sum_{k=1}^Kq_k(s,t;\cdot)\}$, where $P_k(s,t;\cdot),q_k(s,t;\cdot)$ are defined as in \eqref{eq:solution gamma}. Let $e_3=(1,0,0)^\T$, and the online covariance estimate is given by
\begin{equation}
\label{online cov}
\widetilde{\gamma}^{(K)}(s,t)=e_3^\T\left\{\widetilde{P}_{\gamma,1}^{(K)}(s,t)\right\}^{-1}\widetilde{q}_{\gamma,1}^{(K)}(s,t).
\end{equation}

To conclude, we state the following algorithm for the mean estimation.
% , where the calculation of $\eta_{\mu,l}^{(K)}$ and the selection of $\widetilde h_{\mu}^{(K)}$ are given in (\ref{form of eta fda}) of Section \ref{sec:th} and Appendix \ref{appx:band}, respectively. The covariance estimation can be obtained similarly.	
We mention that the proposed method can be applied to the general $d$-dimensional local linear regression as well as higher-order local polynomials. Please see S.3 of Supplementary Material for more detail.

{\small\begin{breakablealgorithm}
\renewcommand{\algorithmicrequire}{\textbf{Input:}}
\renewcommand{\algorithmicensure}{\textbf{Output:}}
\caption{Online Estimation for Mean Function}\label{alg1}
\begin{algorithmic}[1]
% \begin{spacing}{1.35}
%    \REQUIRE $T_{Kij}$ and $Y_{Kij}$, where $i=1,\ldots,n_K,j=1,\ldots,m_Ki$.
% \ENSURE $\widetilde{\mu}^{(K)}(t)$; $\widetilde{P}_{\mu,l}^{(K)}(t)$, $\widetilde{q}_{\mu,l}^{(K)}(t)$, $l=1,2,\ldots,L$.
\STATE \textbf{Initialize:} $K\leftarrow0$, $S_{K,1}\leftarrow0$, $\phi_{\mu,l}^{(K)}\leftarrow0$, $\widetilde{P}_{\mu,l}^{(K)}(t)\leftarrow0$, $\widetilde{q}_{\mu,l}^{(K)}(t)\leftarrow0$, for $l=1,2,\ldots,L$.
\STATE \textbf{While} $K<K_{max}$:
\STATE \quad \textbf{input}: $\{(T_{Kij},Y_{Kij}):i=1,\ldots,n_K,j=1,\ldots,m_{Ki}\}$
\STATE \quad $K\leftarrow K+1$, $s_{K,1}\leftarrow\sum_{i=1}^{n_K}m_{Ki}$, $S_{K,1}\leftarrow S_{K,1}+s_{K,1}$;
\STATE \quad compute $\widetilde{h}_\mu^{(K)}$ as in \eqref{online h} based on pilot estimates as introduced in Appendix \ref{appx:band};
\STATE\quad \textbf{for} $l=1,2,\ldots,L$:
\STATE\quad\quad compute $\eta_{\mu,l}^{(K)}$ based on \eqref{form of eta fda} in Section \ref{sec:th};
\STATE\quad\quad compute $\idx_l^{(K)}$ based on \eqref{eq:I(l)};
\STATE\quad\quad update $\phi_{\mu,l}^{(K)}$ by \eqref{eq:centroids};
\STATE\quad\quad update $\widetilde{P}_{\mu,l}^{(K)}(t)$ and $\widetilde{q}_{\mu,l}^{(K)}(t)$ by \eqref{eq:update Pq}; 
\STATE\quad\textbf{end for}
\STATE\quad \textbf{output}: $\widetilde{\mu}^{(K)}(t)=e_2^\T\{\widetilde{P}_{\mu,1}^{(K)}(t)\}^{-1}\widetilde{q}_{\mu,1}^{(K)}(t)$
\STATE \textbf{end while}
\end{algorithmic}  
\end{breakablealgorithm}}

\section{Theoretical Analysis}
\label{sec:th}

In this section, we illustrate the asymptotic normality of the online mean and covariance estimates of different sampling schemes. Then we study the convergence of online  bandwidth selection and derive the optimal dynamic candidate bandwidth sequence, which gives a lower bound for the relative efficiency versus the batch estimates.

Recall that there are $n_k$ subjects in the $k$th block and the $i$th subject is measured at $m_{ki}$ times for $i=1,\ldots,n_k$, and $s_{K,l},S_{K,l}$ are defined in \eqref{n}. To describe the phase transition phenomenon of the mean and covariance estimators, we denote the total number of subjects up to block $K$ and the individual averages of observations for $\mu$ and $\gamma$ by,
\begin{equation}
\label{N average}
N_K=\sum_{k=1}^Kn_k,\ \overline{m}_{\mu,K}=S_{K,1}/N_K,\ \overline{m}_{\gamma,K}=S_{K,2}/N_K.
\end{equation}
Write $a\asymp b$ if $a/b\rightarrow C\in(0,\infty)$ as $a,b\rightarrow\infty$. Recall that $K$ is the total number of blocks, and $N_K\ge K$ is the number of subjects. We stress that the asymptotic results in this section are for $K\rightarrow \infty$ in spirit of streaming data accumulation.
Recalling that $\omega_{\mu,k}^{(K)}=s_{k,1}/S_{K,1}$ and $\omega_{\gamma,k}^{(K)}=s_{k,2}/S_{K,2}$, we further define 
\begin{equation}
\label{eq:eta moments}
\rho_{\mu,j}^{(K)}=\sum_{k=1}^K\omega_{\mu,k}^{(K)}\left(\widetilde{\eta}_{\mu,k}^{(K)}\right)^j,\ \rho_{\gamma,j}^{(K)}=\sum_{k=1}^K\omega_{\gamma,k}^{(K)}\left(\widetilde{\eta}_{\gamma,k}^{(K)}\right)^j,\ j\in\mathbb{Z}.
\end{equation}
\begin{thm}
\label{THM:MEAN AN}
Under \eqref{assump:obs time point}--\eqref{assump:cov n}, \eqref{assump:mean struct} and \eqref{assump:kernel1}--\eqref{assump:kernel2} in Appendix \ref{appx:assumps}, let $f$ be the density of $T$ and $\rho_{\mu,j}^{(K)}$ be defined as in \eqref{eq:eta moments}.	Suppose that $\widetilde{h}_{\mu}^{(K)}\asymp S_{K,1}^{-1/5}$, $\widetilde{h}_{\mu}^{(K)}=\eta_{\mu,1}^{(K)}>\cdots>\eta_{\mu,L}^{(K)}$, and $\lim\sup_K \overline{m}_{\gamma,K}/\overline{m}_{\mu,K}^2<\infty$, where $\overline{m}_{\mu,K},\overline{m}_{\gamma,K}$ are defined as in \eqref{N average}. Denote $R(W)=\int W(x)^2dx$, $\alpha(W)=\int x^2W(x)dx$ and recall that $\sigma^2$ is the noise variance. For a fixed interior point $t\in(0,1)$, as $K\rightarrow\infty$, the estimate $\widetilde{\mu}^{(K)}(t)$ as in \eqref{online mu} satisfies
\begin{align*}
\left\{S_{K,1}\big/\rho_{\mu,-1}^{(K)}\right\}^{\frac{1}{2}}\left\{\widetilde{\mu}^{(K)}(t)-\mu(t)-\frac{1}{2}\alpha(W)\mu''(t)\rho_{\mu,2}^{(K)}+o_p\left(\rho_{\mu,2}^{(K)}\right)\right\}
\stackrel{d}{\longrightarrow} N\big(0,\Gamma_{\mu}(t)\big),
\end{align*}
where
\begin{equation*}
\Gamma_{\mu}(t)=R(W)\frac{\gamma(t,t)+\sigma^2}{f(t)}+\gamma(t,t)\frac{S_{K,2}}{S_{K,1}\rho_{\mu,-1}^{(K)}},
\end{equation*}
which gives the following phase transition,
\begin{itemize}
\item[(1)] when $\overline{m}_{\mu,K}/N_K^{1/4}\rightarrow0$, 
\begin{align*}
\Gamma_{\mu}(t)=R(W)\frac{\gamma(t,t)+\sigma^2}{f(t)};
\end{align*}
\item[(2)] when $\overline{m}_{\mu,K}/N_K^{1/4}\rightarrow C$, the variance becomes 
\begin{align*}
\Gamma_{\mu}(t)=R(W)\frac{\gamma(t,t)+\sigma^2}{f(t)}+C_1\gamma(t,t),
\end{align*}
where $C_1=\lim_{K}S_{K,2}/\{S_{K,1}\rho_{\mu,-1}^{(K)}\}$, and $C,C_1\in(0,\infty)$;
\item[(3)] when $\overline{m}_{\mu,K}/N_K^{1/4}\rightarrow\infty$, the bias vanishes and the distribution is simplified to 
\begin{align*}
\left\{N_KS_{K,1}^2/S_{K,2}\right\}^{\frac{1}{2}}\Big\{\widetilde{\mu}^{(K)}(t)-\mu(t)\Big\}
\stackrel{d}{\longrightarrow} N\big(0,\gamma(t,t)\big).
\end{align*}
\end{itemize}
\end{thm}

Theorem \ref{THM:MEAN AN} gives a systematic partition of functional data into three categories according to the numbers of repeated measurements on experimental subjects, yielding different convergence properties, which shall be discussed more after representing the asymptotic behaviors of covariance estimate. Note that in case (2) we have $S_{K,1}/S_{K,2}\asymp \overline{m}_{\mu,K}^{-1}\asymp S_{K,1}^{-1/5}$ and $\vert\rho_{\mu,-1}^{(K)}-1/\widetilde{h}_{\mu}^{(K)}\vert=\vert\sum_{k=1}^K\omega_{\mu,k}^{(K)}(1/\widetilde\eta_{\mu,k}^{(K)})-1/\widetilde{h}_{\mu}^{(K)}\vert\le\sum_{k=1}^K\omega_{\mu,k}^{(K)}\vert1/\widetilde{h}_{\mu}^{(k)}-1/\widetilde{h}_{\mu}^{(K)}\vert$, then $S_{K,2}/\{S_{K,1}\rho_{\mu,-1}^{(K)}\}\rightarrow C_1\in(0,\infty)$ holds when $\widetilde{h}_{\mu}^{(K)}\asymp S_{K,1}^{-1/5}$. Though we assume that $T_{kij}$ are i.i.d. across and within subjects, it is noted that when $T_{kij}$ may be dependent within subjects, such as scheduled visit times with random missing and/or fluctuation, the proposed online estimator also attains the same convergence result as the batch estimator by slightly modifying the proofs.

Recall that \cite{zhang2016from} has taken the mean function $\mu$ as known in the covariance estimation, so $\widehat\mu^{(K)}$ is not entangled in the asymptotic distribution of $\widehat\gamma^{(K)}$. In the batch setting, we use the batch estimator $\widehat\mu^{(K)}$ instead and establish the asymptotic normality of the covariance estimation which is proved to be the same as using $\mu$. To see this, denote $e_{kij}^{(K)}=\mu(T_{kij})-\widehat\mu^{(K)}(T_{kij})$ and $\zeta_{kij}=\Phi_{ki}(T_{kij})+\varepsilon_{kij}$, then 
\begin{equation*}
\widehat C_{ki,j_1,j_2}=\left(e_{kij_1}^{(K)}+\zeta_{kij_1}\right)\left(e_{kij_2}^{(K)}+\zeta_{kij_2}\right),\quad C_{ki,j_1,j_2}=\zeta_{kij_1}\zeta_{kij_2}.
\end{equation*}
On one hand, $\var (\widehat{C}_{ki,j_1,j_2})$ and $\var (C_{ki,j_1,j_2})$ are of the same order, both dominated by $\zeta_{kij_1}\zeta_{kij_2}$. On the other, note that $\widehat{\mu}^{(K)}$ is the weighted average of $\{Y_{kij}:1\le k \le K,1\le i\le n_k,1\le j\le m_{ki}\}$, and $e_{ki,j_1}^{(K)}$ is weakly correlated to $\zeta_{kij_2}$, which makes the bias induced by $e_{ki,j_1}^{(K)}\zeta_{kij_2}$ negligible. The bias of $e_{ki,j_1}^{(K)}e_{ki,j_2}^{(K)}\asymp(\widehat h_\mu^{(K)})^4$ is also negligible compared to $(\widehat h_\gamma^{(K)})^2$.
This argument also applies to the online estimation, and indicates that the bandwidth selection for mean and covariance estimators can be decoupled, which greatly facilitates the derivation. For space economy, we present the result for the online covariance estimation (using $\widetilde\mu^{(K)}$), and defer the batch version (using $\widehat\mu^{(K)}$) to Lemma \ref{lem:cov an} in Appendix \ref{appx:assumps}, while the detailed proofs are offered in the Supplementary Material. Denote
{\small\begin{align}
\label{V1 V2 V3}
V_1(s,t)&=\var\big\{\big(Y_1-\mu(T_1)\big)\big(Y_2-\mu(T_2)\big)|T_1=s,T_2=t\big\},\nonumber\\
V_2(s,t)&=\cov\big\{\big(Y_1-\mu(T_1)\big)\big(Y_2-\mu(T_2)\big),\big(Y_1-\mu(T_1)\big)\big(Y_3-\mu(T_3)\big)|T_1=s,T_2=t,T_3=t\big\},\\
V_3(s,t)&=\cov\big\{\big(Y_1-\mu(T_1)\big)\big(Y_2-\mu(T_2)\big),\big(Y_3-\mu(T_3)\big)\big(Y_4-\mu(T_4)\big)|T_1=s,T_2=t,T_3=s,T_4=t\big\}.\nonumber
\end{align}}	
\begin{thm}
\label{THM:COV AN}
Under \eqref{assump:obs time point}--\eqref{assump:cov n} and \eqref{assump:cov struct}--\eqref{assump:kernel2} in Appendix \ref{appx:assumps}, let $\rho_{\gamma,j}^{(K)}$ be defined as in \eqref{eq:eta moments}. Suppose that $\widetilde{h}_{\gamma}^{(K)}\asymp S_{K,2}^{-1/6}$, $\widetilde{h}_{\gamma}^{(K)}=\eta_{\gamma,1}^{(K)}>\cdots>\eta_{\gamma,L}^{(K)}$, and $f,R(W),\alpha(W),\sigma^2$ are the same as in Theorem \ref{THM:MEAN AN}. 
For a fixed interior point $(s,t)\in(0,1)^2$, as $K\rightarrow\infty$, the covariance estimator $\widetilde{\gamma}^{(K)}$ in \eqref{online cov} satisfies
{\small\begin{align*}
\left\{S_{K,2}\big/\rho_{\gamma,-2}^{(K)}\right\}^{\frac{1}{2}}\left\{\widetilde{\gamma}^{(K)}(s,t)-\gamma(s,t)-\frac{1}{2}\alpha(W)\left(\frac{\partial^2\gamma}{\partial s^2}+\frac{\partial^2\gamma}{\partial t^2}\right)\rho_{\gamma,2}^{(K)}+o_p\left(\rho_{\gamma,2}^{(K)}\right)\right\}\stackrel{d}{\longrightarrow} N\big(0,\Gamma_{\gamma}(s,t)\big),
\end{align*}}
where
\begin{align*}
\Gamma_{\gamma}(s,t)=&\{1+I(s=t)\}\left\{R(W)^2\frac{V_1(s,t)}{f(s)f(t)}\right.\\
&\left.+\frac{1}{S_{K,2}\rho_{\gamma,-2}^{(K)}}\sum_{k=1}^K\frac{s_{k,3}}{\widetilde\eta_{\gamma,k}^{(K)}}R(W)\frac{f(s)V_2(t,s)+f(t)V_2(s,t)}{f(s)f(t)}\right\}+\frac{S_{K,4}}{S_{K,2}\rho_{\gamma,-2}^{(K)}}V_3(s,t),
\end{align*}
and the following statements hold,
\begin{itemize}
\item[(1)] when $\overline{m}_{\gamma,K}/N_K^{1/2}\rightarrow0$,
\begin{align*}
\Gamma_{\gamma}(s,t)=\{1+I(s=t)\}\frac{R(W)^2V_1(s,t)}{f(s)f(t)};
\end{align*}
\item[(2)] when $\overline{m}_{\gamma,K}/N_K^{1/2}\rightarrow C$,
\begin{align*}
\Gamma_{\gamma}(s,t)=&\{1+I(s=t)\}\left\{\frac{R(W)^2V_1(s,t)}{f(s)f(t)}\right.\\
&\left.+\frac{R(W)C_1}{C_0^{1/2}}\frac{f(s)V_2(t,s)+f(t)V_2(s,t)}{f(s)f(t)}\right\}+C_1^2V_3(s,t),
\end{align*}
where $C_0=\lim_K S_{K,2}S_{K,4}/S_{K,3}^2$, $C_1^2=\lim_KS_{K,4}/\{S_{K,2}\rho_{\gamma,-2}^{(K)}\}$, and $C,C_0,C_1\in(0,\infty)$;
\item[(3)] when $\overline{m}_{\gamma,K}/N_K^{1/2}\rightarrow\infty$, the bias vanishes and the distribution is simplified to 
\begin{align*}
\left\{N_KS_{K,2}^2\big/S_{K,4}\right\}^{\frac{1}{2}}\Big\{\widetilde{\gamma}^{(K)}(s,t)-\gamma(s,t)\Big\}\stackrel{d}{\longrightarrow} N\big(0,V_3(s,t)\big).
\end{align*}
\end{itemize}
\end{thm}

Similar to case (2) of Theorem \ref{THM:MEAN AN}, $C_1^2=\lim_KS_{K,4}/\{S_{K,2}\rho_{\gamma,-2}^{(K)}\}\in(0,\infty)$ holds as long as $\widetilde h_{\gamma}^{(K)}\asymp S_{K,2}^{-1/6}$. Note that the partition in Theorem \ref{THM:COV AN} is identical to that in Theorem \ref{THM:MEAN AN}. Case (1) in these two theorems is the so-called {\em sparse} design, where mean and covariance estimates attain nonparametric convergence rate that is slower than root-$N_K$. Case (2) and (3) are {\em dense} data such that the parametric convergence rate $N_K^{1/2}$ can be achieved. We refer to case (2) as {\em moderately dense} and case (3) {\em ultra dense} for that, under the latter scheme, both the bias term and the variance discontinuity vanish. Such phase transition phenomenon for online estimation is accordant with the classical batch result in \cite{zhang2016from}. Using the same techniques, one can further prove that, in terms of $L_2$ and uniform convergences, the online estimates can also attain the rates as the batch results in \cite{zhang2016from}. 

To derive the optimal bandwidths, we introduce the integrated mean squared error. For a function $m(x)$  and its estimate $\widehat m(x)$, let $f_X$ be the density function of $X$ supported on $I$, then $IMSE(\widehat m)$ is a global loss criterion defined as
\begin{equation*}
IMSE(\widehat m)=\int_I\left[\mean\{\widehat m(x)-m(x)\}^2\right]f_X(x)dx.
\end{equation*}
Note that $\mean\{\widehat m(x)-m(x)\}^2=\var\widehat m(x)+[\mean\{\widehat m(x)\}-m(x)]^2$, we have the following results for the batch and online estimates.
\begin{cor}\label{COR:IMSE BATCH}
Under \eqref{assump:obs time point}--\eqref{assump:cov n} and \eqref{assump:mean struct}--\eqref{assump:kernel2} in Appendix \ref{appx:assumps}, as $K\rightarrow\infty$,
\begin{align*}
IMSE\left(\widehat\mu^{(K)}\right)
&=\frac{1}{4}\theta_\mu\alpha^2(W)\big\{\widehat h_\mu^{(K)}\big\}^4+\frac{\nu_\mu}{S_{K,1}\widehat h_\mu^{(K)}}+o_p\left(\big\{\widehat h_\mu^{(K)}\big\}^4+S_{K,1}^{-1}\big\{\widehat h_\mu^{(K)}\big\}^{-1}\right),\\
IMSE\left(\widehat\gamma^{(K)}\right)
&=\frac{1}{4}\theta_\gamma\alpha^2(W)\big\{\widehat h_\gamma^{(K)}\big\}^4+\frac{\nu_\gamma}{S_{K,2}\big\{\widehat h_\gamma^{(K)}\big\}^2}+o_p\left(\big\{\widehat h_\gamma^{(K)}\big\}^4+S_{K,2}^{-1}\big\{\widehat h_\gamma^{(K)}\big\}^{-2}\right),
\end{align*}
where 
\begin{align}
\label{theta nu}
&\theta_{\mu}=\int \mu''(t)f(t)dt,\ \theta_{\gamma}=\int\int(\partial^2\gamma/\partial s^2+\partial^2\gamma/\partial t^2)f(s)f(t)dsdt,\nonumber\\
&\nu_{\mu}=\int\Gamma_{\mu}(t)f(t)dt,\ \nu_{\gamma}=\int\int\Gamma_{\gamma}(s,t)f(s)f(t)dsdt.
\end{align}
\end{cor}
\begin{cor}\label{COR:IMSE ONLINE}
Suppose that the conditions in Theorem \ref{THM:MEAN AN}--\ref{THM:COV AN} hold, then as $K\rightarrow\infty$,
\begin{align*}
IMSE\left(\widetilde\mu^{(K)}\right)
&=\frac{1}{4}\theta_\mu\alpha^2(W)\big\{\rho_{\mu,2}^{(K)}\big\}^2+\frac{\nu_\mu}{S_{K,1}}\rho_{\mu,-1}^{(K)}+o_p\left(\left\{\rho_{\mu,2}^{(K)}\right\}^2+S_{K,1}^{-1}\rho_{\mu,-1}^{(K)}\right),\\
IMSE\left(\widetilde\gamma^{(K)}\right)
&=\frac{1}{4}\theta_\gamma\alpha^2(W)\big\{\rho_{\gamma,2}^{(K)}\big\}^2+\frac{\nu_\gamma}{S_{K,2}}\rho_{\gamma,-2}^{(K)}+o_p\left(\left\{\rho_{\gamma,2}^{(K)}\right\}^2+S_{K,2}^{-1}\rho_{\gamma,-2}^{(K)}\right),
\end{align*}
where $\theta_{\mu},\theta_{\gamma},\nu_{\mu},\nu_{\gamma}$ are defined as in Corollary \ref{COR:IMSE BATCH}. 
\end{cor}     
From Corollary \ref{COR:IMSE BATCH}, the optimal bandwidths to minimize $IMSE\left(\widehat\mu^{(K)}\right)$ and $IMSE\left(\widehat\gamma^{(K)}\right)$ are, respectively,
\begin{equation}
\label{opt hs}
h_{\mu,*}^{(K)}=\left(\frac{\nu_\mu}{\alpha^2(W)\theta_\mu}\right)^\frac{1}{5}S_{K,1}^{-\frac{1}{5}},\quad h_{\gamma,*}^{(K)}=\left(\frac{\nu_\gamma}{\alpha^2(W)\theta_\gamma}\right)^\frac{1}{6}S_{K,2}^{-\frac{1}{6}}.
\end{equation}
We suggest to use pilot estimates adopting the online method to approximate the unknown integrals $\theta_\mu,\theta_\gamma$ and the variance terms $\nu_\mu,\nu_\gamma$, see details in Appendix \ref{appx:band}. The online estimates of optimal bandwidths are
{\small\begin{equation}
\label{online h}
\widetilde{h}_{\mu}^{(K)}=\left(\frac{\widetilde\nu_{\mu}^{(K)}}{\alpha^2(W)\widetilde\theta_{\mu}^{(K)}}\right)^\frac{1}{5}S_{K,1}^{-\frac{1}{5}},\ \widetilde{h}_{\gamma}^{(K)}=\left(\frac{\widetilde\nu_{\gamma}^{(K)}}{\alpha^2(W)\widetilde\theta_{\gamma}^{(K)}}\right)^\frac{1}{6}S_{K,2}^{-\frac{1}{6}},
\end{equation}}
where $\widetilde\theta_\mu^{(K)},\widetilde\theta_\gamma^{(K)}$ and $\widetilde\nu_\mu^{(K)},\widetilde\nu_\gamma^{(K)}$ are given explicitly in Appendix \ref{appx:band}. 
{Let $\widehat h_\mu^{(K)}$ be the bandwidth selected by replacing the online pilot estimates in \eqref{online h} with the batch ones.} When the bandwidths for pilot estimates are of appropriate orders, the online bandwidths \eqref{online h} can attain the same convergence rates as the batch competitors, as stated in the following theorem.

\begin{thm}
\label{THM:BAND}
Suppose that the conditions in Theorem \ref{THM:MEAN AN}--\ref{THM:COV AN} hold. 
Let $\widetilde{h}_{\mu}^{(K)}$ and $\widetilde{h}_{\gamma}^{(K)}$ be the proposed online bandwidth estimates \eqref{online h} where $\widetilde\theta_\mu^{(K)},\widetilde\theta_\gamma^{(K)}$ and $\widetilde\nu_\mu^{(K)},\widetilde\nu_\gamma^{(K)}$ are based on online local polynomials with bandwidths and candidates given explicitly in \eqref{h theta nu fda} of Appendix \ref{appx:band}.
Then the online bandwidth selection satisfies that as $K\rightarrow\infty$,
\begin{equation*}
\frac{\widetilde{h}_{\mu}^{(K)}-h_{\mu,*}^{(K)}}{h_{\mu,*}^{(K)}}=O_p\left(S_{K,1}^{-\frac{2}{7}}\right),\quad
\frac{\widetilde{h}_{\gamma}^{(K)}-h_{\gamma,*}^{(K)}}{h_{\gamma,*}^{(K)}}=O_p\left(S_{K,2}^{-\frac{1}{4}}\right),
\end{equation*}
where $h_{\mu,*}^{(K)},h_{\gamma,*}^{(K)}$ are the optimal bandwidths \eqref{opt hs}.
\end{thm}

Define the relative efficiency in terms of integrated mean squared errors by
\begin{equation*}
eff_K(\widetilde{\mu}^{(K)})=IMSE(\widetilde{\mu}^{(K)})/IMSE(\widehat{\mu}^{(K)}), \quad
eff_K(\widetilde{\gamma}^{(K)})=IMSE(\widetilde{\gamma}^{(K)})/IMSE(\widehat{\gamma}^{(K)}),
\end{equation*}
which measures the performance of the proposed online method compared to the classical estimate using full data. We now present our key result, the choice of candidate bandwidth sequence which makes $\widetilde{\eta}_{\mu,k}^{(K)}$ (or $\widetilde{\eta}_{\gamma,k}^{(K)}$) as close to $\widetilde{h}_\mu^{(K)}$ (or $\widetilde{h}_\gamma^{(K)}$) as possible for all $k$ and maximizes the relative efficiency with a lower bound.

\begin{thm}
\label{THM:EFF}
Under \eqref{assump:obs time point}--\eqref{assump:kernel2} in Appendix \ref{appx:assumps}, when the candidate bandwidths are
\begin{equation}
\label{form of eta fda}
\eta_{\mu,l}^{(K)}=\left(\frac{L-l+1}{L}\right)^{\frac{1}{5}}\widetilde{h}_{\mu}^{(K)},\quad \eta_{\gamma,l}^{(K)}=\left(\frac{L-l+1}{L}\right)^{\frac{1}{6}}\widetilde{h}_{\gamma}^{(K)},\quad l=1, \ldots, L,
\end{equation} 
then as $K\rightarrow\infty$, the corresponding online estimates $\widetilde{\mu}^{(K)}$ and $\widetilde{\gamma}^{(K)}$ attain the optimal relative efficiency with the following lower bound,
\begin{equation*}
\left\{1+\left(\frac{2d}{d+4}c_1+\frac{4}{d+4}c_2\right)\frac{1}{L}+\frac{d}{d+4}c_2^2\frac{1}{L^2}\right\}^{-1}+O_p\left(S_{K,d}^{-\frac{2}{d+6}}\right),
\end{equation*}
where $d=1$ for $\widetilde{\mu}^{(K)}$ and $d=2$ for $\widetilde\gamma^{(K)}$. Specifically,
\begin{align} \label{eq:lb}
& eff(\widetilde{\mu}^{(K)})\geq(1+0.1831L^{-1}+0.0032L^{-2})^{-1}+O_p\left(S_{K,1}^{-2/7}\right),\nonumber\\
& eff(\widetilde{\gamma}^{(K)})\geq(1+0.2422L^{-1}+0.0190L^{-2})^{-1}+O_p\left(S_{K,2}^{-1/4}\right).
\end{align}
\end{thm}

\begin{figure}[htbp]
\centering
\includegraphics[width = 5.8in]{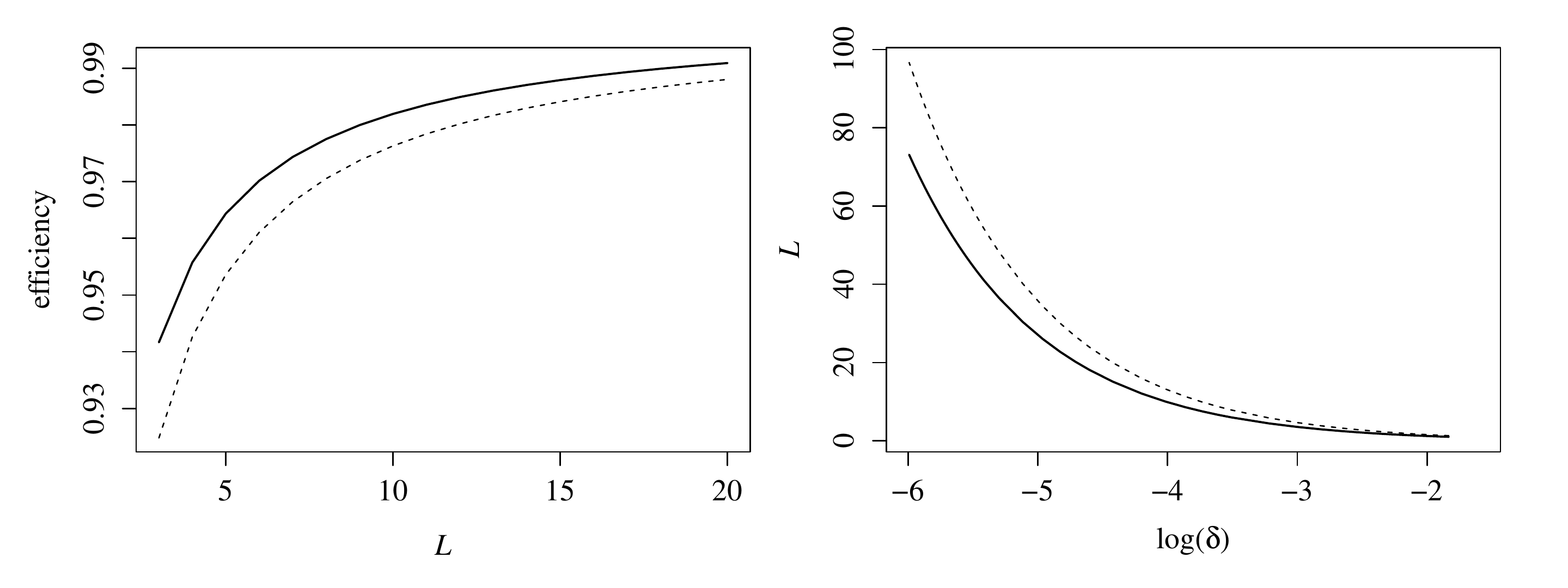}
\caption{\label{fig:theoretical eff} The left panel shows lower bounds for the relative efficiencies of the proposed online mean (solid lines) and covariance (dashed lines) estimates versus different lengths $L$ of candidate bandwidth sequences. The right panel plots the selection of $L$ under different upper bounds of $\log(\delta)$ for the proposed online mean (solid lines) and covariance (dashed lines) estimates.}
\end{figure}

From the illustration in the left panel of Figure \ref{fig:theoretical eff}, the relative efficiencies of the proposed online mean and covariance estimates increase rapidly to exceed 95\% when $L\geq5$. This is desirable to attain high efficiency with a small/moderate $L$. Note that the computational cost in terms of time and memory is proportional to $L$, this lower bound helps make an informed trade-off between statistical and computational efficiency, which makes the proposed method practically useful. We also define the relative error,
\begin{equation*}
\delta(\widetilde{\mu}^{(K)})=1-eff(\widetilde{\mu}^{(K)}),\quad\delta(\widetilde{\gamma}^{(K)})=1-eff(\widetilde{\gamma}^{(K)}),
\end{equation*}
whose upper bound can be derived from \eqref{eq:lb}. We plot $L$ versus the upper bound of $\log(\delta)$ in the right panel of Figure \ref{fig:theoretical eff} as a suggestion for the selection of $L$ in practice.

We conclude this section by mentioning that the asymptotic distributions of $\widetilde{\mu}^{(K)}$ and $\widetilde{\gamma}^{(K)}$ have the consistent form with the general $d$-dimensional local linear regression. Hence Theorem \ref{THM:BAND} and  \ref{THM:EFF} hold for general local linear regression with $d$-dimensional covariates. See details in S.4 of Supplementary Material.
% \nolinenumbers

\section{Simulation}
\label{sec:simu}

We conduct simulation to illustrate the performance of the proposed online method and verify the theoretical findings in Section \ref{sec:th}. Let noises $\v_{kij}$ be i.i.d. from $N(0,0.5^2)$ and $T_{kij}$ follow a uniform distribution on $[0,1]$. Data are generated by $Y_{kij}=X_{ki}(T_{kij})+\v_{kij}$, where $X_{ki}(T_{kij})=\mu(T_{kij})+\Phi_{ki}(T_{kij})$ with the mean function $\mu(t)=2\sin(2\pi t)$ and the stochastic part $\Phi(t)=\sum_{i=1}^{10}\xi_i\phi_i(t)$, where $\phi_1(t)=1$, $\phi_i(t)=\sqrt{2}\cos\{(i-1)\pi t\}$, $\xi_i$ are independently sampled from $N(0,\lambda_i)$ with $\lambda_i=0.4\times i^{-2},\ i=1,\ldots,10$. Recall that there are $n_k$ subjects in the $k$th block, among which the $j$th one has $m_{ki}$ observations. For sparse data, we let $n_k$ follow the normal distribution $N(20,9)$ and $m_{ki}$ follow $N(6,4)$; for dense data, we let $n_k=3$ and $m_{ki}$ follow $N(20,4)$; which are all rounded off to the nearest integers.  The experiment is repeated 100 times, each with $K_{max}=1000$ blocks.

We derive in \eqref{h theta nu fda} of Appendix \ref{appx:band} that bandwidths for pilot estimates of $\theta_\mu,\theta_\gamma$ and $\nu_\mu,\nu_\gamma$ in \eqref{theta nu} shall be
\begin{equation*}
h_{\theta_\mu}^{(K)}=GS_{K,1}^{-1/7},\ h_{\nu_\mu}^{(K)}=RS_{K,1}^{-1/5},\ 
h_{\theta_\gamma}^{(K)}=GS_{K,2}^{-1/8},\ h_{\nu_\gamma}^{(K)}=RS_{K,2}^{-1/6}. 
\end{equation*}
After experimenting an extensive range of $G$ and $R$, we find that the convergence rate of the online bandwidth is not sensitive to the values of these two quantities. Values in $[0.5^{1/d},1]$ are in general adequate, thus we set $G=R=0.5^{1/d}$ in the sequel, where $d=1$ for estimating $\mu$ and $d=2$ for estimating $\gamma$. Let $J$ be the length of candidate bandwidth sequences for the pilot estimates of $\theta_\mu,\theta_\gamma$ and $\nu_\mu,\nu_\gamma$, and recall that $L$ is the length of candidate bandwidth sequences for estimating $\mu$ and $\gamma$, we set $J=L$ when estimating $\mu$ and $J=3$ when estimating $\gamma$ to ease computation. We also stop updating $\widetilde{\theta}_{\gamma}^{(K)}$ and $\widetilde{\nu}_{\gamma}^{(K)}$ after $K'=200$ to further reduce computation without influencing the convergence rate as $K'/K_{max}=O_p(1)$. Note that the classical batch method is computational expensive when sample size tends large, we implement it at every 40 blocks, i.e. $K=40,80,120,\ldots$. For mean and covariance estimation of sparse and dense data, we examine the following measures. 

\begin{figure}[htbp]
\centering
\includegraphics[width = 6.5in,height=1.8in]{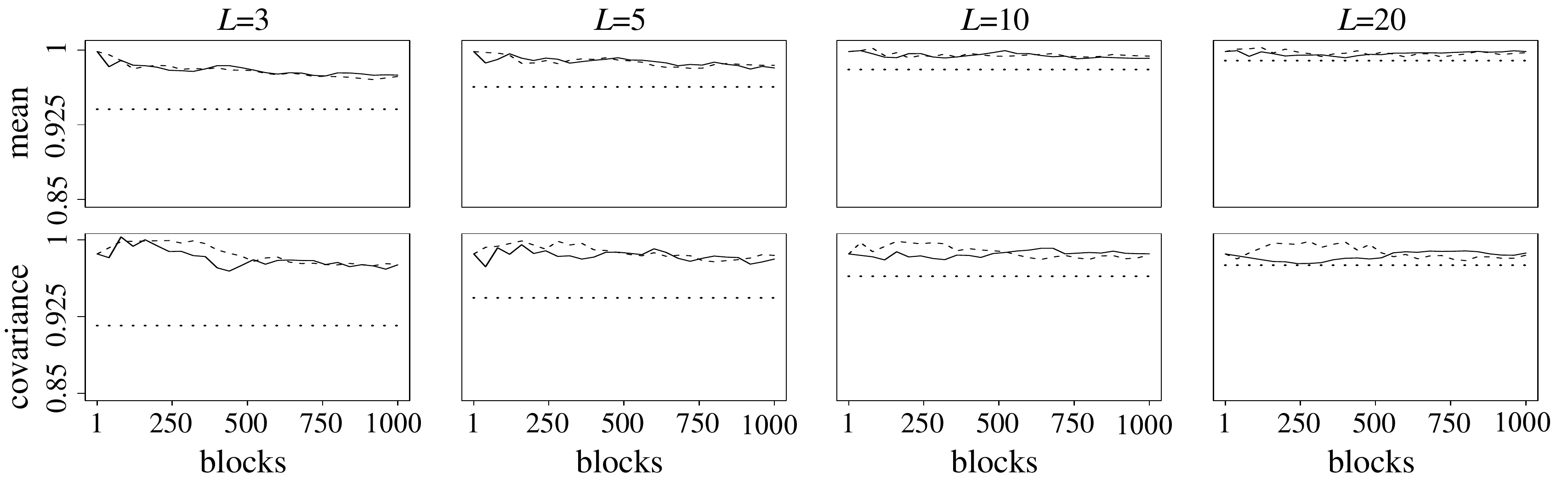}
\caption{\label{fig:empirical eff} The empirical relative efficiencies of the proposed online mean and covariance estimates for functional data based on sparse design (solid) and dense design (dashed) versus the theoretical lower bound (dotted) with candidate bandwidth sequence length $L=3,5,10$ and $20$.}
\end{figure}

\textit{1. Relative efficiency.} Figure \ref{fig:empirical eff} shows the empirical relative efficiencies of the mean and covariance estimates that increase with $L$ and are stably higher than the theoretical lower bounds in Theorem \ref{THM:EFF} when $K$ tends large.	Recall that \cite{kong2019efficiency} studied the $d$-dimensional online local linear regression of independent noises, which is not applicable to functional data. Even under the assumption of independent errors, their theoretical lower bounds of relative efficiency, which are 0.943 when $d=1$ and 0.924 when $d=2$, are still smaller than ours. An explanation is that their stored statistics are fixed, and if the bandwidths of previous blocks deviate from the optimal values, there is no data-driven adjustment in the sequential estimates. By contrast, our stored statistics are updated dynamically by selecting different bandwidths from the candidate sequences and hence produce more efficient estimates.

\begin{figure}[htbp]
\centering
\includegraphics[width = 6.5in,height=2.2in]{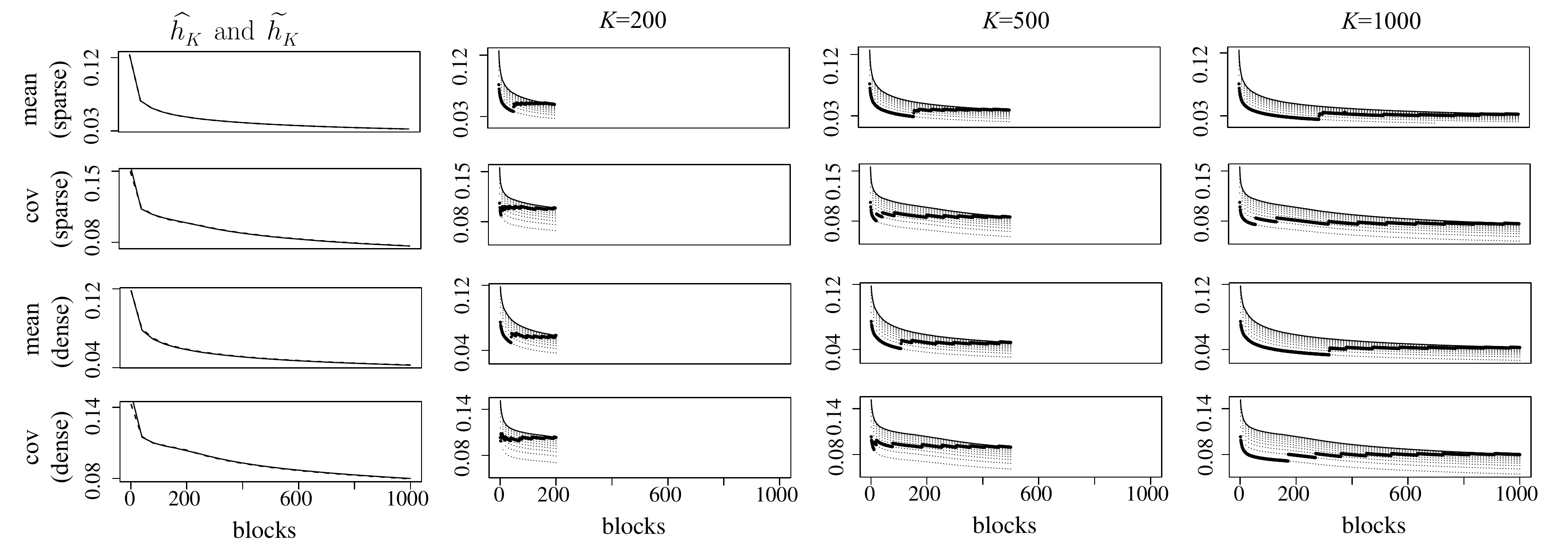}
\caption{\label{fig:band} The four rows show, respectively, the bandwidth selection for mean and covariance estimation of sparse and dense data. The first column shows the Monto Carlo averages of the bandwidths selected by the online method (solid) $\widetilde{h}_{\mu}^{(K)},\widetilde{h}_{\gamma}^{(K)}$ and the batch method $\widehat{h}_{\mu}^{(K)},\widehat{h}_{\gamma}^{(K)}$ (dashed) based on 100 runs for $K=1,2,\ldots,1000$, and the other three columns depict the dynamically updated bandwidths $\{\widetilde{\eta}_{\mu,k}^{(K)}\}_{k=1}^K,\{\widetilde{\eta}_{\gamma,k}^{(K)}\}_{k=1}^K$ (thick dots) at time $K=200,500,1000$, respectively, along with the dynamic candidate sequences (light dots) and optimal bandwidth estimates (connected by solid line) at each $k$.}
\end{figure}

\textit{2. Bandwidth selection.} The convergence of bandwidths in Theorem \ref{THM:BAND} is also examined in Figure \ref{fig:band}, where both batch and online selections converge to the theoretical optimal bandwidth along with data collection. 
Moreover, we depict the dynamically updated pseudo-bandwidth sequences $\{\widetilde{\eta}_{\mu,k}^{(K)}\}_{k=1}^K$ and $\{\widetilde{\eta}_{\gamma,k}^{(K)}\}_{k=1}^K$ that are  used to produce the estimates at time $K=200,500,1000$, similar to that shown in Figure \ref{fig:sketch}. This provides empirical support for the fact that the our method is indeed able to adjust the sub-statistics in previous blocks in spite of no access to those data.

\begin{figure}[htbp]
\centering
\includegraphics[width = 6.5in,height=1.7in]{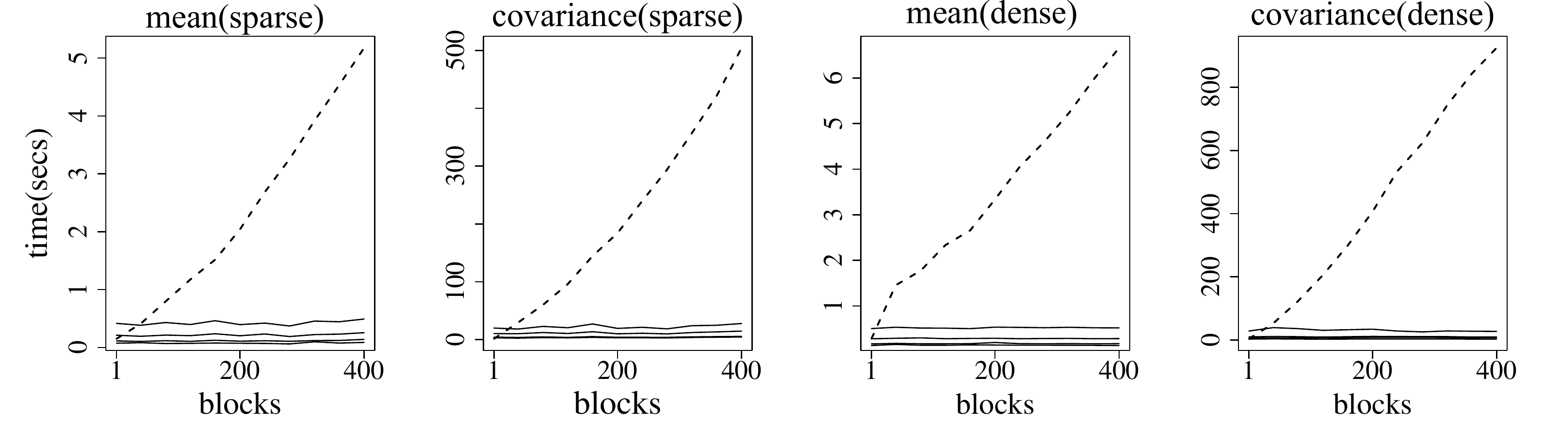}
\caption{\label{fig:simutime} The comparison of computing times between the proposed online method and the batch method when estimating mean and covariance functions based on sparse and dense functional data. The dashed lines correspond to the computing times of the batch method and the solid lines to the online method, from the lowest to the highest in each panel representing $L=3,5,10, 20$, respectively.}
\end{figure}

\textit{3. Computational time.} We compare the computing times of the classical batch and the proposed online methods using the Unix server of 2.10GHz CPU and 188G memory with 176 logic cores. It is noted from Figure \ref{fig:simutime} that the computing time of the classical batch method grows approximately linearly with data blocks, while the online method spends nearly constant time given similar block sizes. Thus we graphed only the first 400 blocks for visualization when compared with the online method using $L=3,5,10, 20$ that achieves substantial computational saving proportional to $L$. 

We close this section by suggesting a reasonable range of $L\in[5, 20]$ based on our empirical and theoretical findings. The choice shall be made depending on whether accuracy or computation is of main concern.

\section{Real Data Examples}
\label{sec:real data}

In this section, we present two real data examples to illustrate the usefulness of the proposed online method.

\subsection{Airline delay example}
The airline dataset consists of flight arrival and departure details for all commercial airports in the USA, from January 1989 to December 2000 (\url{https://community.amstat.org/jointscsg-section/dataexpo/dataexpo2009}). This involves 304 airports whose daily numbers of flights range from 1 to 1269. We are interested in the pattern of delay time (unit: minute) during peak hours, i.e., from 6:00 to 23:00. Airports with flights less than 50 per day are removed. For each airport, flights departed during off-peak hours are also removed. Suppose that the departure delay time is independent across days and we treat the daily airport as a subject with flight departure delay times as measurements. Let one block represent a day and there are 4383 blocks in total.
We randomly select $n_k$ airports from the $k$th day to form the $k$th data block, where $n_k$ are discretely distributed with equal probability among $\{6,7,\ldots,15\}$. To obtain comparable results, the dense and sparse designs use the same airports for each block, and the dense measurements contain the sparse ones for each subject. Recall the number of measurements of the $j$th subject in the $k$th block is denoted by $m_{ki}$.	For dense scheme, we let $m_{ki}$ follow the discrete uniform distribution on $\{15,16,\ldots,20\}$ and randomly pick $m_{ki}$ flights with equal probability as measurements; for sparse case, choose $m_{ki}$ equal-likely from 8 to 10 to form the measurements, where $j=1,\ldots, n_k$, $k=1,\ldots,4383$. We also plot the results obtained from all airports and all flights as baseline for comparison. 
% In other words, denote the data used for the sparse, dense and baseline estimates by $\mathcal{D}_s$, $\mathcal{D}_d$ and $\mathcal{D}_b$, and we have $\mathcal{D}_s\subset\mathcal{D}_d\subset\mathcal{D}_b$.

\begin{figure}[htbp]
\centering
\includegraphics[width = 5.8in,height=2.2in]{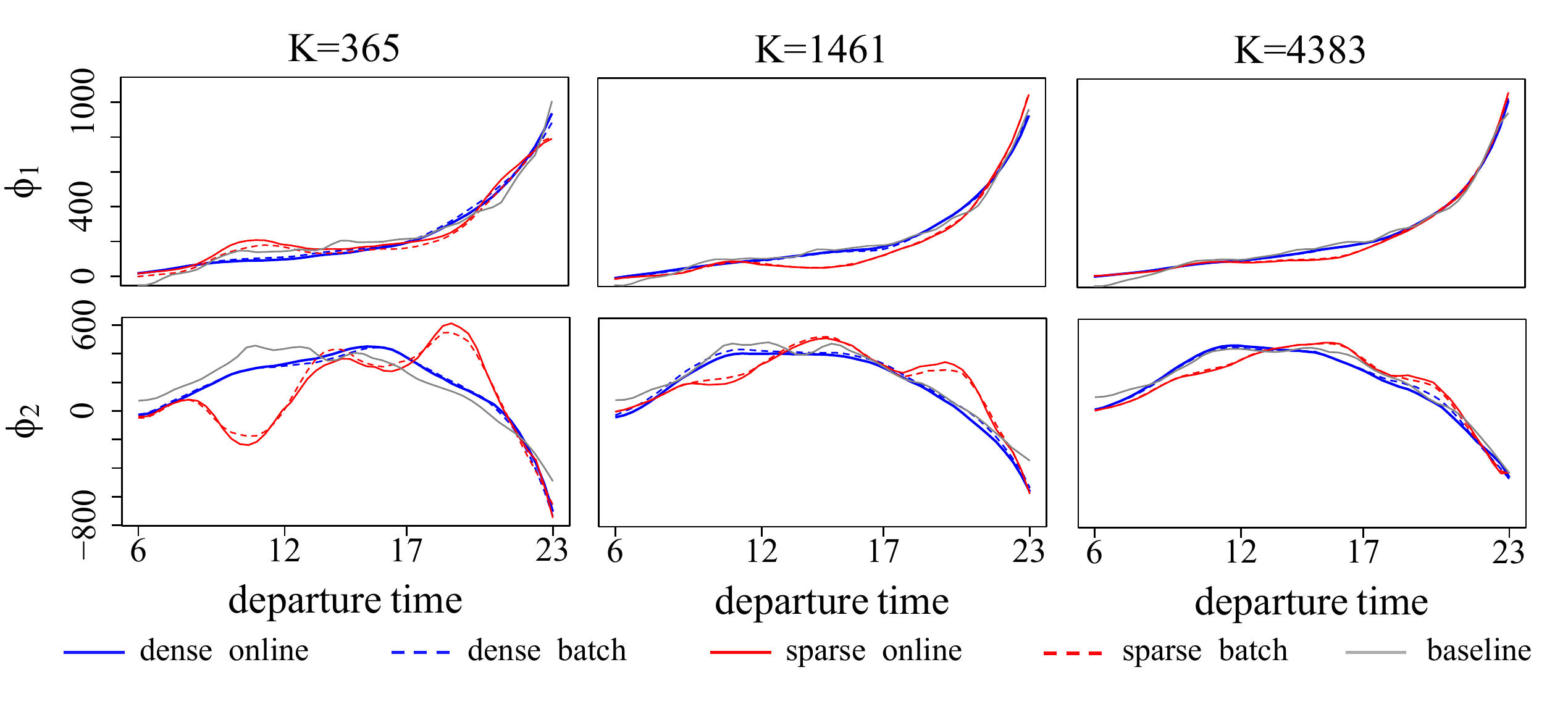}
\caption{\label{fig:flights_vecs}\small Estimates of the first two eigenfunctions $\phi_1,\phi_2$ for the airline delay dataset at $K = 365, 1461, 4383$ (the end of year 1989, 1992, 2000), where the dashed lines correspond to the batch results and the solid to the proposed online method with dense design plotted in blue and sparse in red. The first eigenfunction $\phi_1$ has similar pattern to the mean function, and occupies 88\% FVE. The estimates based on all airports and flights are also plotted in gray as the baseline.}
\end{figure}

\begin{figure}[htbp]
\centering
\includegraphics[width = 6.5in]{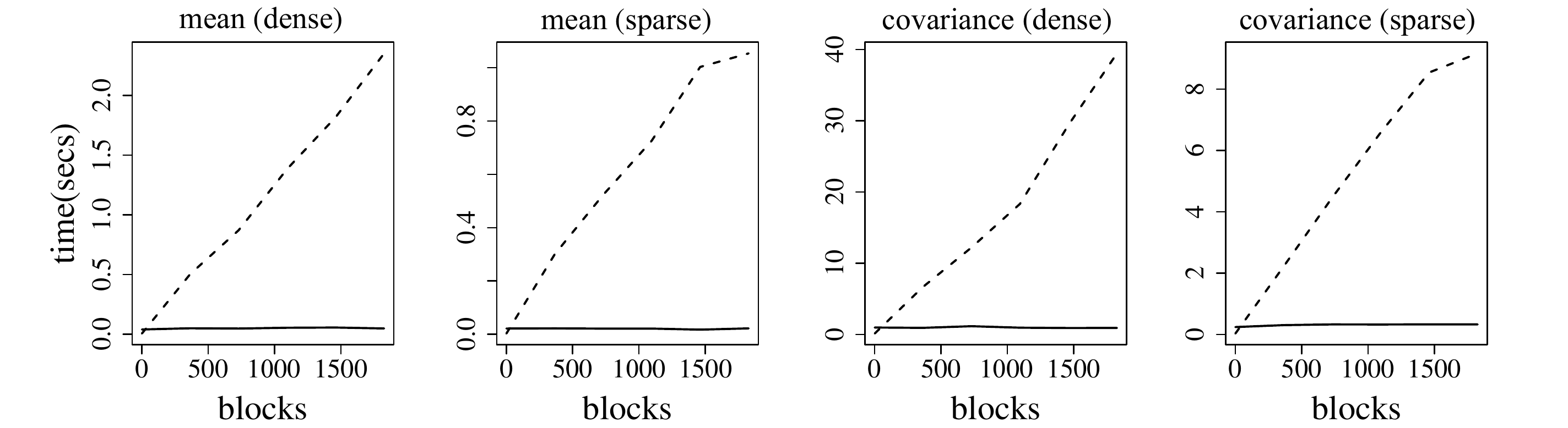}
\caption{\label{fig:flights_time} The comparison of computing times for the airline delay dataset between our method and the batch method of the first 1821 blocks/five years for mean and covariance estimation. The dashed lines correspond to the time complexity of the batch method and the solid lines to our method. }
\end{figure}

We set $L = J = 10$ for the estimation of mean function, and $L = 10$ and $J = 3$ for covariance estimation, where $L$ is the length of candidate bandwidth sequence for main estimation, and $J$ is for the pilot bandwidths. As in simulation, we update the pilot estimates in bandwidth selection for covariance estimation till $K = 200$. Delay aggravates along time, which is probably due to the cumulative effect of congestion for flight runways. For both designs, the online and the batch estimates agree well when more data enter the model. We also conduct functional principal component analysis based on the covariance estimates. We show the first two eigenfunctions in Figure \ref{fig:flights_vecs} at time $K = 365, 1461, 4383$ (the end of year 1989, 1992, 2000), whose aggregated fractions of variation explained (FVE) exceeds 95\%. 
% Due to the differences of the data used ($\mathcal{D}_s$, $\mathcal{D}_d$ and $\mathcal{D}_b$), the patterns of baseline, dense and sparse estimates show slight differences at the beginning, but the distinctions narrow down as data accumulate. 
In both dense and sparse designs, as data accumulate, the online estimates become closer to the batch ones as well as the baseline results using all airports and flights. The first component has similar pattern to the mean function showing more variation in later hours, while the second represents the influence of different handling strategies of airports. An explanation may be that those airports with sound management system and adequate resource matching can alleviate the aggravating effect of delay to a certain extent, and vice versa.  Comparison of computation times is given in Figure \ref{fig:flights_time}, where the online method shows substantial gains over the batch method.

\begin{figure}[htbp]
\centering
\includegraphics[width = 6in,height=3.3in]{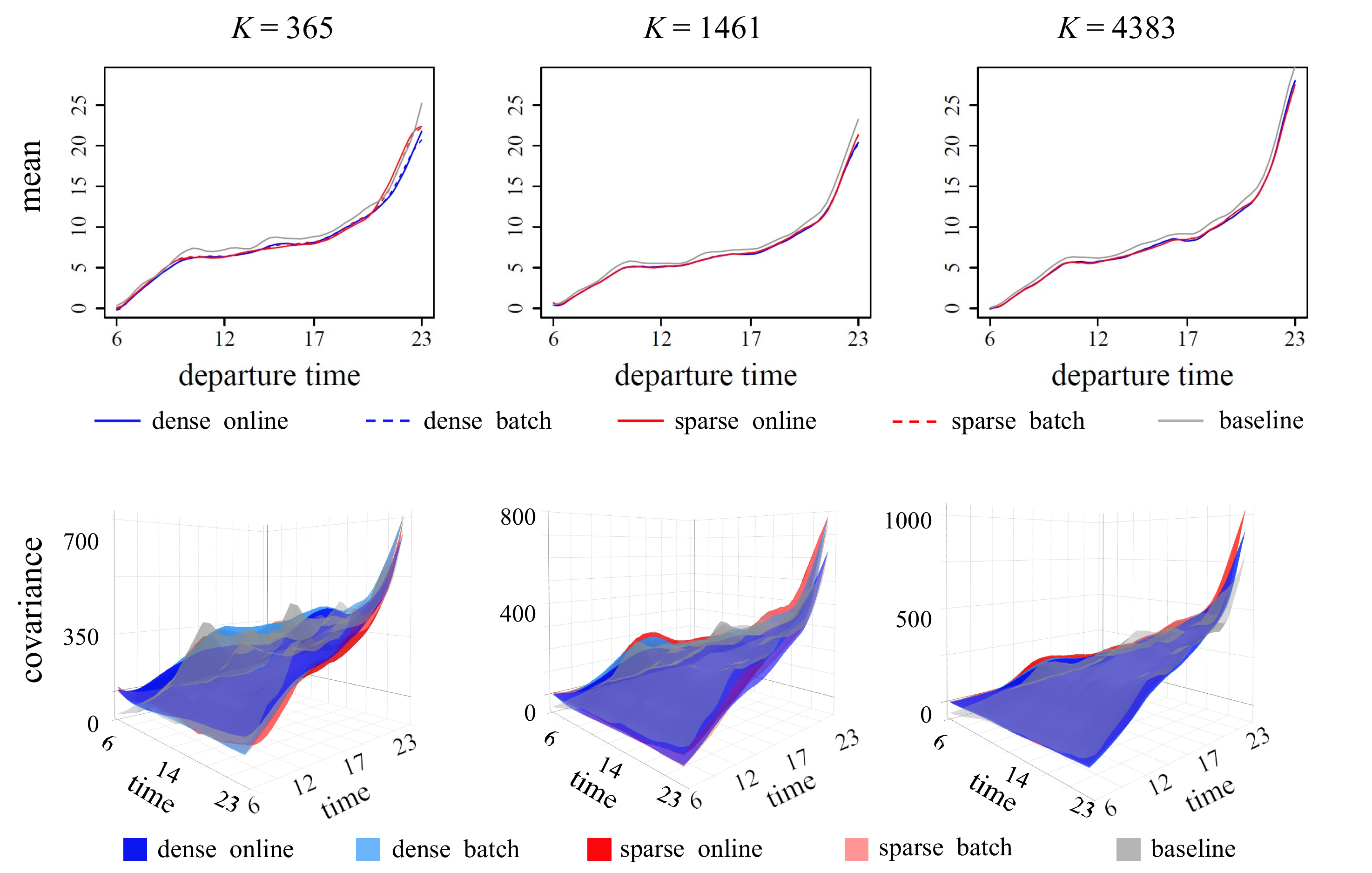}
\caption{\label{fig:flights_mean}\small Mean and covariance estimates for the airline delay dataset at $K = 365, 1461, 4383$ (the end of year 1989,1992,2000), where the dashed lines and light-colored surfaces represent the batch results, and the solid lines and dark-colored surfaces represent the proposed online estimates, with color blue representing the dense case and red representing the sparse case. The estimates based on all airports and flights are also plotted in gray as the baseline.	}
\end{figure}

\subsection{Online news example}

The second example is the online news data, which contains 93239 news items published during 2015/11/08 to 2016/07/07 and their hourly social feedback (number of clicks) of 48 hours after publishing on LinkedIn, Facebook and Google, which can be downloaded from \url{http://archive.ics.uci.edu/ml/datasets/News+Popularity+in+Multiple+Social+Media+Platforms?tdsourcetag=s_pctim_aiomsg}. Denote the publish time for the $j$th news as $t_{0,j}$ and measurement time as $t_{ij}=t_{0,j}+i\Delta t$, where $i=1,\ldots,48$, $t_{0,j}\in(0,24]$ and $\Delta t=1$ (unit: hour). The release time $t_{0,j}$ can be viewed as random, and hence the observations are irregular by space. We refer to the news published during 0:00 to 6:00 as morning news and aim to study their number of clicks per hour over the domain $[6,30]$. i.e., 24 hours after 6:00. We add up the number of clicks from the four platforms for the morning news and further remove the inactive news whose total number of clicks is less than 6. The remaining 5689 news are divided into 1138 blocks. We set the blockwise number of subjects as $n_k=5$ for $k\le1137$ and $n_k=4$ for $k=1138$. For each subject, let the dense measurements contain the sparse ones. For dense scheme, we set $m_{ki}=24$ and for sparse scheme $m_{ki}=5$ with measurements randomly chosen from the dense set with equal probability, $j=1,\ldots,n_k$, $k=1,\ldots,1138$. Let $Y_{kij}^0$ denote the original hourly click number, and we make the following transformation: $Y_{kij}=\log(Y_{kij}^0+1)$.

\begin{figure}[htbp]
\centering
\includegraphics[width = 5.5in,height=3.2in]{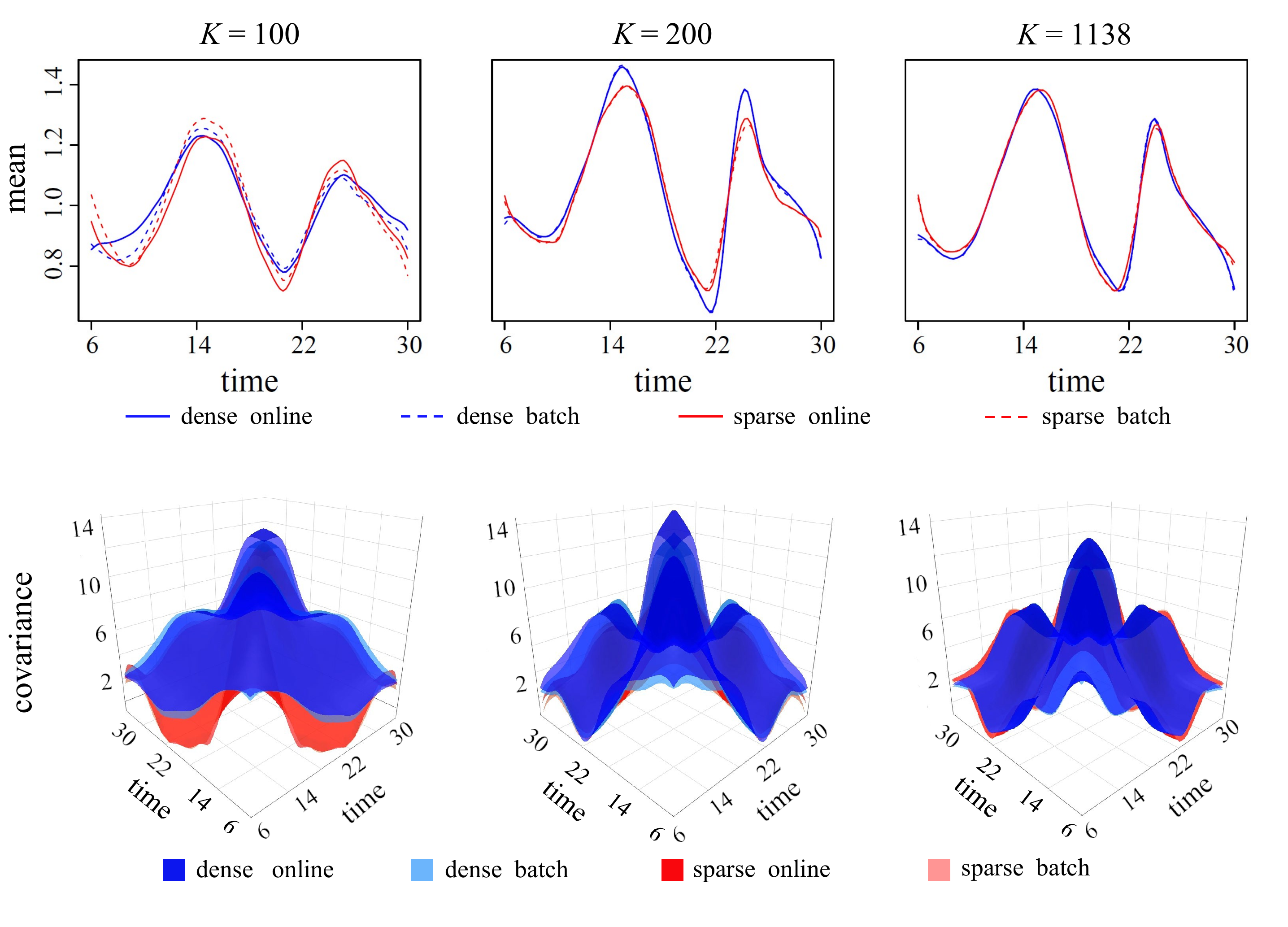}
\caption{\label{fig:news_mean}\small Mean and covariance estimates for the online news dataset when $K=100,200,1138$, where the dashed lines and light-colored surfaces correspond to the batch results, and the solid lines and light-colored surfaces to the online estimates, with color blue representing the dense case and red representing the sparse case. The first eigenfunction $\phi_1$ has similar pattern to the mean function, and occupies nearly 66\% FVE.}
\end{figure}

\begin{figure}[htbp]
\centering
\includegraphics[width = 5.5in,height=2.8in]{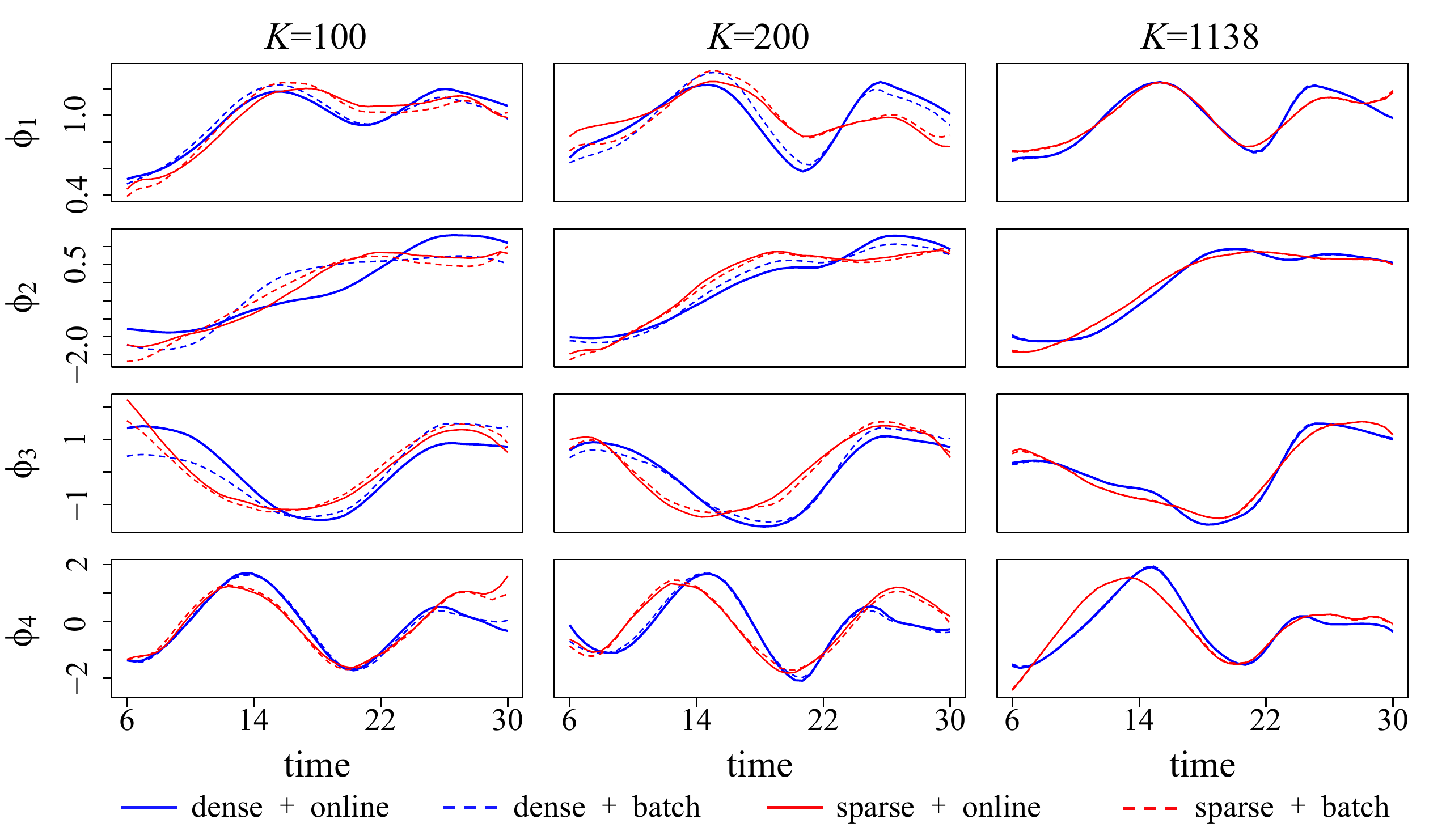}
\caption{\label{fig:news_vecs} Estimates of the first four eigenfunctions $\phi_1,\ldots,\phi_4$ for the online news dataset using the online (solid) and batch (dashed) methods at time $K=100,200,1138$ with dense design plotted in blue and sparse in red.}
\end{figure}

\begin{figure}[htbp]
\centering
\includegraphics[width = 6.5in]{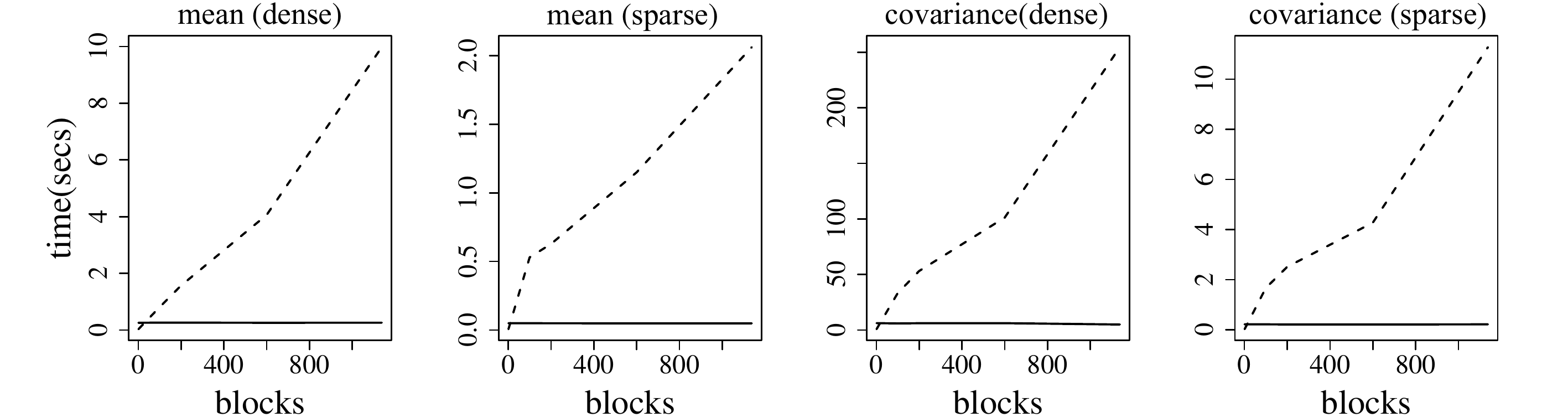}
\caption{\label{fig:news_time}\small The comparison of computing times for the online news dataset between our method and the batch method from $K=1$ to $K=1138$ for mean and covariance estimation. The dashed lines correspond to the time complexity of the batch method and the solid lines to our method.}
\end{figure}

Parameter $L$ and $J$ are set the same as in the airline delay data and the pilot estimates in bandwidth for covariance estimation update till $K = 200$. Estimates for the mean and covariance function, and eigenfunctions are shown in Figure \ref{fig:news_mean}-\ref{fig:news_vecs}, where difference between online estimates and batch ones narrows down when data accumulate. Mean function estimates in Figure \ref{fig:news_mean} reveals that the hourly clicks attain peaks at 14:00 and 00:00, i.e., after the lunch break and before night sleeping. Here we plot the first four eigenfunction estimates whose aggregated FVE is about 95\%. The first eigenfunction indicates again that the attention a news received is influenced by people's daily activity pattern. The second represents sustained growth or decrease of the news consumption, which is determined by the news quality, as people like to share valuable news, while articles that attract hits by headlines are gradually ignored. Finally, one can conclude from Figure \ref{fig:news_time} that our online method saves substantial computational time compared to the batch method.

\section{Concluding Remarks}

In this work, we propose a dynamic candidate bandwidth method and apply it to functional data analysis for the mean and covariance estimation in the online context. By pre-storing a sequence of statistics and selecting appropriate candidates, our method dynamically updates statistics for all blocks. To the best of our knowledge, we are the first to implement dynamic update of previous statistics with changing bandwidths for streaming data. The proposed method is both computational and statistical efficient. It requires nearly constant memory and computing time and has a lower bound for the relative efficiency in terms of integrated mean squared errors. The computation and estimation trade-off is achieved by the parameter $L$ that provides useful guidance in practice.

The proposed method can be adopted to the higher-order local polynomials as well as other kernel-type estimates. For instance, one can combine the dynamic candidate bandwidths with profiling algorithm to fit semiparametric partially linear models, or with backfitting algorithm for additive models. Due to the interaction between component estimates in the iterative algorithm, the pseudo-sufficient statistics for each component function would depend on the estimates of other components and the relative efficiency needs to be further investigated. 

\section*{Supplementary Material}
The supplementary material contains the proofs for Theorem 1-4 and Lemma 1.

\section*{Appendix}
\begin{appendix}
\renewcommand{\theenumi}{\Alph{section}.\arabic{enumi}}
\renewcommand{\labelenumi}{(\theenumi)}
\renewcommand{\theequation}{\Alph{section}.\arabic{equation}}
\section{Assumptions and Auxiliary results}\label{appx:assumps}

Recall that we observe $Y_{kij}=X_{ki}(T_{kij})+\v_{kij}$, where $X_{ki}(T_{kij})=\mu(T_{kij})+\Phi_{ki}(T_{kij})$, $j=1,\ldots,m_{ki},i=1,\ldots,n_{k},k=1,\ldots,K$, with $\cov\{X(s),X(t)\}=\gamma(s,t)$. Denote $\overline{m}_{j,K}=S_{K,j}/N_K\textrm{ for }j\ge3$. The following assumptions are imposed for theoretical analysis in Section \ref{sec:th}.

\begin{enumerate}
\item\label{assump:obs time point} The observed time points $\{T_{kij}\}$ are i.i.d. copies of a random variable $T$ defined on [0,1]. The density $f(\cdot)$ of $T$ is bounded away from 0 and its second derivative is bounded.
\item\label{assump:indpt} $X$ is independent of $T$, and $\v$ is independent of $T$ and $\Phi$.
\item\label{assump:deri} The second and fourth derivatives of $\mu$ and $\gamma$ are bounded and continuous.
\item\label{assump:cov n} The sample size satisfies $\lim_K\overline{m}_{\gamma,K}\cdot\overline{m}_{4,K}/\overline{m}_{3,K}^2=C_0\in[1,\infty]$, $\lim\sup_K\overline{m}_{\gamma,K}/\overline{m}_{\mu,K}^2$, $\lim\sup_K\overline{m}_{\mu,K}\cdot\overline{m}_{3,K}/\overline{m}_{\gamma,K}^2$, $\lim\sup_K\overline{m}_{4,K}/\overline{m}_{\gamma,K}^2<\infty$, where $\overline{m}_{l,K},\overline{m}_{\mu,K}, \overline{m}_{\gamma,K}$ are defined in \eqref{N average}.
\item\label{assump:block} $\max_{k\le K} s_{k,j} /S_{K,j}\rightarrow 0$ as $K\rightarrow\infty$ for $j=1,2$.
\item\label{assump:mean struct} $\sup_{t\in[0,1]}\mean|\Phi(t)|^3<\infty$ and $\mean|\v|^3<\infty$.
\end{enumerate}
Assumption \eqref{assump:obs time point}--\eqref{assump:cov n} are general for functional mean and covariance estimation, see \cite{zhang2016from}. Assumption \eqref{assump:block} is reasonable to require the number of measurements in each block is relatively small compared to the total number of measurements in all $K$ blocks. Assumption \eqref{assump:mean struct} is a standard requirement for mean estimation. For covariance estimation, further assumption on the moments of the random part is needed.
\begin{enumerate}
\setcounter{enumi}{6}
\item\label{assump:cov struct} $\sup_{t\in[0,1]}\mean|\Phi(t)|^6<\infty$ and $\mean|\v|^6<\infty$.
\end{enumerate}
We also impose the following assumptions on the kernel function. 
\begin{enumerate}
\setcounter{enumi}{7}
\item\label{assump:kernel1} $W(\cdot)$ is a symmetric probability density function on $[-1,1]$ and $\alpha(W),R(W)<\infty$.
% \begin{equation*}
% \alpha(W)=\int u^2W(u)du<\infty,\ \  R(W)=\int W(u)^2du<\infty.
% \end{equation*}
\item\label{assump:kernel2} $W(\cdot)$ is Lipschitz continuous: There exists $0<L<\infty$ such that
\begin{equation*}
|W(u)-W(v)|\leq L|u-v|,\ \  {\rm for\ any\ } u,v\in[0,1].
\end{equation*}
\end{enumerate}

The lemma below states the asymptotic normality of the batch covariance estimate based on the batch estimator $\widehat{\mu}^{(K)}$, corresponding to Theorem \ref{THM:COV AN}, which also improves the result of  \cite{zhang2016from} that used the underlying mean function $\mu$.
\begin{lem}
\label{lem:cov an}
Under \eqref{assump:obs time point}--\eqref{assump:cov n} and \eqref{assump:cov struct}--\eqref{assump:kernel2} in Appendix \ref{appx:assumps}, suppose that $\widehat h_{\gamma}^{(K)}-h_{\gamma,*}^{(K)}=o_p(N_K^{-1/6})$, and let $f$ be the density of $T$. 
Denote $R(W)=\int W(x)^2dx$, $\alpha(W)=\int x^2W(x)dx$ and recall that $\sigma^2$ is the noise variance. 
For a fixed interior point $(s,t)\in(0,1)^2$, as $K\rightarrow\infty$, the covariance estimator $\widehat{\gamma}^{(K)}$ in \eqref{eq:solution gamma} satisfies
{\footnotesize\begin{align*}
\left\{S_{K,2}\left(\widehat h_{\gamma}^{(K)}\right)^{2}\right\}^{\frac{1}{2}}\left\{\widehat{\gamma}^{(K)}(s,t)-\gamma(s,t)-\frac{1}{2}\alpha(W)\left(\frac{\partial^2\gamma}{\partial s^2}+\frac{\partial^2\gamma}{\partial t^2}\right)\left(\widehat h_{\gamma}^{(K)}\right)^2+o_p\left(\left(\widehat h_{\gamma}^{(K)}\right)^2\right)\right\}\stackrel{d}{\longrightarrow} N\big(0,\Gamma_{\gamma}'(s,t)\big),
\end{align*}}
where
{\footnotesize\begin{align*}
\Gamma_{\gamma}'=&\{1+I(s=t)\}\left\{S_{K,2}^{-1}R(W)^2\frac{V_1(s,t)}{f(s)f(t)}\right.\\
&\left.+\widehat h_{\gamma}^{(K)}S_{K,2}^{-2}S_{K,3}R(W)\frac{f(s)V_2(t,s)+f(t)V_2(s,t)}{f(s)f(t)}\right\}+V_3(s,t)\left(\widehat h_{\gamma}^{(K)}\right)^{-2}S_{K,4}/S_{K,2}^2,
\end{align*}}
\end{lem}
The proof for the above lemma is similar to the online case which can be found in S.2 of the Supplementary Material. 

\renewcommand{\theenumi}{\alph{enumi}}
\renewcommand{\labelenumi}{(\theenumi)}
\section{Bandwidth Selection}
\label{appx:band}

Recall that the optimal bandwidths for estimating $\mu$ and $\gamma$ are, respectively,
\begin{eqnarray*}
h_{\mu,*}^{(K)}=\left(\frac{\nu_\mu}{\alpha^2(W)\theta_\mu}\right)^\frac{1}{5}S_{K,1}^{-\frac{1}{5}},\quad 
h_{\gamma,*}^{(K)}=\left(\frac{\nu_\gamma}{\alpha^2(W)\theta_\gamma}\right)^\frac{1}{6}S_{K,2}^{-\frac{1}{6}},
\end{eqnarray*}
where $\theta_{\mu}=\int \mu''(t)f(t)dt$, $\theta_{\gamma}=\int\int(\partial^2\gamma/\partial s^2+\partial^2\gamma/\partial t^2)f(s)f(t)dsdt$, $\nu_{\mu}=\int\Gamma_{\mu}(t)f(t)dt$ and $\nu_{\gamma}=\int\int\Gamma_{\gamma}(s,t)f(s)f(t)dsdt$, where $\Gamma_\mu,\Gamma_\gamma$ are delineated below.
We suggest to use pilot estimates adopting online approach to approximate these quantities.
Take $\theta_\mu$ as an example. We first estimate $\mu''$ by an online local cubic smoother with candidate bandwidth sequence $\{\eta_{\theta_\mu,j}^{(k)}\}_{j=1}^J$, $k=1,2,\ldots,K$, and use $\widetilde{f}^{(K)}=e_2^\T\widetilde{P}_{\theta_\mu,1}^{(K)}e_2$ to estimate $f$, then integrate $(\widetilde{\mu}'')^{(K)}\widetilde{f}_K$ to obtain $\widetilde{\theta}_\mu^{(K)}$. And $\theta_\nu$ can be estimated by the same method, with candidate sequence denoted by $\{\eta_{\theta_\gamma,j}^{(k)}\}_{j=1}^J$.
When estimating $\nu_\mu$ and $\nu_\gamma$, the density function can be estimated by $e_2^\T\widetilde{P}_{\nu_\mu,1}^{(K)}e_2$ and $e_3^\T\widetilde{P}_{\nu_\gamma,1}^{(K)}e_3$, and we need only estimate $\Gamma_\mu$ and $\Gamma_\gamma$, which have distinguished expressions in the cases listed below and require different estimation procedures. These estimates are based on other local polynomials whose bandwidth selection is specified after presenting the methodology.

\begin{enumerate}
\item\label{sparse mean} \textit{Mean function estimation for sparse data.} Recall the definitions of $\gamma(t,t)$ and $\sigma^2$ in the underlying models are \eqref{model fda}, and
\begin{align*}
\Gamma_{\mu}=R(W)\frac{\gamma(t,t)+\sigma^2}{f(t)}.
\end{align*}
We need only estimate $r(t)=\gamma(t,t)+\sigma^2$.
We follow the method of \cite{fan1998variance} to estimate $r(t)$: 
\begin{itemize}
\item[1.] Estimate $\mu(t)$ by an online local linear smoother denoted by $\check{\mu}^{(K)}$ with bandwidth $h_{\nu_\mu}^{(K)}=O_p(S_{K,1}^{-1/5})$ and the corresponding candidate bandwidth sequence $\{\eta_{\nu_\mu,j}^{(K)}\}_{j=1}^J$, $k=1,2,\ldots,K$; 
\item[2.] Compute $\widehat{r}_{Kij}=\{Y_{Kij}-\check{\mu}^{(K)}(T_{Kij})\}^2$, and smooth $\widehat{r}_{Kij}$ by an online local linear smoother with the above $h_{\nu_\mu}^{(K)}$ and $\{\eta_{\nu_\mu,j}^{(K)}\}_{j=1}^J$ to obtain $\widetilde{r}^{(K)}(t)$.
\end{itemize}

\item\label{sparse cov} \textit{Covariance function estimation for sparse data.} Recall the definition of $V_1$ in \eqref{V1 V2 V3} and 
\begin{align*}
\Gamma_{\gamma}=\{1+I(s=t)\}\frac{R(W)^2V_1(s,t)}{f(s)f(t)}.
\end{align*}
We only need estimate $V_1(s,t)$ that equals
$V_1(s,t)=\mean\Phi(s)^2\Phi(t)^2+\sigma^2\{\gamma(s,s)+\gamma(t,t)\}+\sigma^4-\gamma(s,t)^2$,
where $\sigma^2$ is the noise variance and $\Phi$ is the stochastic part of $X$. Estimating $V_1(s,t)$ consists of three parts:
\begin{itemize}
\item[1.] Estimate $\gamma(s,t)$ by an online local linear smoother with bandwidth $h_{\nu_\gamma}^{(K)}=O_p(S_{K,2}^{-1/6})$ and the corresponding candidate bandwidth sequence $\{\eta_{\nu_\gamma,j}^{(k)}\}_{j=1}^J$, $k=1,2,\ldots,K$. Denote the estimate by $\check{\gamma}^{(K)}(s,t)$; 
\item[2.] Estimate $\sigma^2$ by $\check{\sigma}_K^2(t)=\widetilde{r}^{(K)}(t)-\check{\gamma}^{(K)}(t,t)$, where $\widetilde{r}^{(K)}(t)$ is obtained  when estimating $\mu$ in step 2 of \eqref{sparse mean}, and note that it is assumed that the noise variance is a constant, we estimate $\sigma^2$ by $\check{\sigma}_K^2=\int\check{\sigma}_K^2(t)dt$;
\item[3.] Smooth $\{\widehat{C}_{Ki,j_1,j_2}-\check{\gamma}^{(K)}(T_{Kij_1},T_{Kij_2})\}^2$ by an online local linear smoother with the above $h_{\nu_\gamma}^{(K)}$ and $\{\eta_{\nu_\gamma,j}^{(K)}\}_{j=1}^J$ to obtain the estimate $\check{\mean}^{(K)}\Phi(s)^2\Phi(t)^2$.
\end{itemize}
Based on these, 
{\small$\widetilde{V}_{1}^{(K)}(s,t)=\check{\mean}^{(K)}\Phi(s)^2\Phi(t)^2+\check{\sigma}_K^2\{\check{\gamma}^{(K)}(s,s)+\check{\gamma}^{(K)}(t,t)\}+\check{\sigma}_K^4-\check{\gamma}^{(K)}(s,t)^2$}.

\item\label{dense mean} \textit{Mean function estimation for dense data.} Recall that for dense data, 
\begin{eqnarray*}
\Gamma_{\mu}=R(W)r(t)/f(t)+\gamma(t,t)
\end{eqnarray*}
Estimating $\gamma$ based on the raw covariance is infeasible as $\mu$ is not available.
As in \eqref{sparse mean}, $r(t)$ is estimated by $\widetilde{r}^{(K)}(t)$, we can make use of the dense measurements for each subject $(k,i)$, pre-smooth $\{Y_{kij}\}_{j=1}^{m_{ki}}$ to obtain an estimate of $X_{ki}(t)+\Phi_{ki}(t)$ which gives the estimate $\widetilde{\sigma}_K^2$ of noise variance $\sigma^2$ \citep{zhang2007inference} and estimate $\gamma(t,t)$ by $\widetilde{r}^{(K)}(t)-\widetilde{\sigma}_K^2$:
\begin{itemize}
\item[1.] Adopt the same method of (a) to compute $\widetilde{r}^{(K)}(t)$;
\item[2.] Pre-smooth to estimate $\sigma^2$. Specifically, for each $j$ at block $K$, apply local linear smooth to $\{Y_{Kij}:j=1,\ldots,m_{Ki}\}$ with bandwidth $h=O_p\big(\overline{m}_{\mu,K}^{-1/5}\big)$ to obtain smoothed $\widehat{Y}_{Kij}$,
then $\check{\v}_{Kij}=Y_{Kij}-\widehat{Y}_{Kij}$. To de-bias, let $\widehat{\v}_{Kij}=\check{\v}_{Kij}-\sum_{j=1}^{m_{Ki}}\check{\v}_{Kij}/m_{Ki}$
and estimate $\sigma^2$ by 
$$\widetilde{\sigma}_K^2=\frac{\sum_{i=1}^{n_K}\sum_{j=1}^{m_{Ki}}\widehat{\v}_{kij}^2}{S_{K,1}}+\frac{S_{K-1,1}}{S_{K,1}}\widetilde{\sigma}_{K-1}^2,$$
then estimate $\gamma(t,t)$ by $\widetilde{r}^{(K)}(t)-\widetilde{\sigma}_K^2$.
\end{itemize}

\item\label{dense cov} \textit{Covariance function estimation for dense data.} Recall that
{\small\begin{align*}
\Gamma_{\gamma}=&\{1+I(s=t)\}\left\{\frac{R(W)^2V_1(s,t)}{f(s)f(t)}+\frac{R(W)C_1}{C_0^{1/2}}\frac{f(s)V_2(t,s)+f(t)V_2(s,t)}{f(s)f(t)}\right\}+C_1^2V_3(s,t),
\end{align*}}
where $V_1,V_2,V_3$ are defined as in \eqref{V1 V2 V3} of Section \ref{sec:th} and can be written as $V_1(s,t)=\mean\Phi(s)^2\Phi(t)^2+\sigma^2\{\gamma(s,s)+\gamma(t,t)\}+\sigma^4-\gamma(s,t)^2$, $V_2(s,t)=\mean\Phi(s)^2\Phi(t)^2+\sigma^2\gamma(t,t)-\gamma(s,t)^2$, $V_3(s,t)=\mean\Phi(s)^2\Phi(t)^2-\gamma(s,t)^2$.
% \begin{align*}
% &V_1(s,t)=\mean\Phi(s)^2\Phi(t)^2+\sigma^2\{\gamma(s,s)+\gamma(t,t)\}+\sigma^4-\gamma(s,t)^2,\\
% &V_2(s,t)=\mean\Phi(s)^2\Phi(t)^2+\sigma^2\gamma(t,t)-\gamma(s,t)^2,\\
% &V_3(s,t)=\mean\Phi(s)^2\Phi(t)^2-\gamma(s,t)^2.
% \end{align*}
One can solve $C_1$ by $C_1=\{\nu_\gamma/(\alpha^2(W)\theta_\gamma)\}^{1/6}$ based on approximations of $V_1,V_2$ and $V_3$.
The estimation of $V_1(s,t), V_2(s,t)$ and $V_3(s,t)$ consists of the three parts below:
\begin{itemize}
\item[1.] Adopt the same method as in step 1 of (b) to compute $\check{\gamma}^{(K)}(s,t)$; 
\item[2.] Estimate $\sigma^2$ by $\widetilde{\sigma}_K^2$ in step 2 of (c);
\item[3.] Adopt the method in step 3 of (b) to compute $\check{\gamma}^{(K)}(s,t)$.
\end{itemize}
\end{enumerate}

The above processes to estimate $\theta_\mu,\theta_\gamma$ and $\nu_\mu,\nu_\gamma$ involve further bandwidth selection. Take the standard local linear regression as an instance.
According to \cite{wand1994kernel} and \cite{fan1998variance}, the optimal bandwidths for estimating $\theta_\mu,\theta_\gamma$ and $\nu_\mu,\nu_\gamma$ are 
{\small\begin{equation*}
h_{\theta_\mu}^{(K)}=G_\mu^*S_{K,1}^{-1/7},\ h_{\theta_\gamma}^{(K)}=G_\gamma^*S_{K,2}^{-1/8},\ h_{\nu_\mu}^{(K)}=R_\mu^*S_{K,1}^{-1/5},\ h_{\nu_\gamma}^{(K)}=R_\gamma^*S_{K,2}^{-1/6}, 
\end{equation*}}
where  $G_\mu^*,G_\gamma^*,R_\mu^*,R_\gamma^*$ involve unknown quantities depending on $\mu$ an $\gamma$. As literature on classical bandwidth selection has demonstrated, the convergences of $\widetilde{h}_\mu^{(K)},\widetilde{h}_\gamma^{(K)}$ in \eqref{online h} hold as long as the bandwidths for estimating $\theta_\mu,\theta_\gamma,\nu_\mu$ and $\nu_\gamma$ satisfy, for $G_\mu,G_\gamma,R_\mu,R_\gamma=O_p(1)$,
{\small\begin{eqnarray}
\label{h theta nu fda}
&h_{\theta_\mu}^{(K)}=G_\mu S_{K,1}^{-1/7},\ h_{\theta_\gamma}^{(K)}=G_\gamma S_{K,2}^{-1/8},\ h_{\nu_\mu}^{(K)}=R_\mu S_{K,1}^{-1/5},\ h_{\nu_\gamma}^{(K)}=R_\gamma S_{K,2}^{-1/6}.%, \nonumber\\
% &G_\mu=O_p(1),\ G_\gamma=O_p(1),\ R_\mu=O_p(1),\ R_\gamma=O_p(1).
\end{eqnarray}}
Hence it is usually adequate to select appropriate constants to compute $h_{\theta_\mu}^{(K)},h_{\theta_\gamma}^{(K)},h_{\nu_\mu}^{(K)},h_{\nu_\gamma}^{(K)}$. Based on extensive numerical experiments, we recommend to set $G_\mu,R_\mu\in[0.5,1]$ and $G_\gamma,R_\gamma\in[0.7,1]$.
Using the same argument of deriving \eqref{form of eta fda}, we use the following candidate bandwidths for $h_{\theta}^{(K)}$ and $h_{\nu}^{(K)}$, for $j=1,2,\ldots,J$,
{\small\begin{align}
\label{eta theta nu fda}
&\eta_{\theta_\mu,j}^{(K)}=\{(J-j+1)/J\}^{1/7}h_{\theta_\mu}^{(K)},\ 
\eta_{\nu_\mu,j}^{(K)}=\{(J-j+1)/J\}^{1/5}h_{\nu_\mu}^{(K)},\nonumber\\
&\eta_{\theta_\gamma,j}^{(K)}=\{(J-j+1)/J\}^{1/8}h_{\theta_\gamma}^{(K)},\ 
\eta_{\nu_\gamma,j}^{(K)}=\{(J-j+1)/J\}^{1/6}h_{\nu_\gamma}^{(K)}.
\end{align}}
So far we have delineated the whole process of online estimation for mean and covariance functions in functional data. 

\end{appendix}

\small
\bibliographystyle{jasa}
\bibliography{paper-ref}

\end{document}